\pdfoutput=1
\documentclass[10pt,letterpaper]{article}
\usepackage{opex3b}
\usepackage{amssymb}
\usepackage{amsmath}
\usepackage{cite}

\begin{document}

%%%%%%%%%%%%%%%%%% title page information %%%%%%%%%%%%%%%%%%
\title{\bf Temporal mode selectivity by frequency conversion in second-order nonlinear optical waveguides} 

\author{D. V. Reddy,$^1$ M. G. Raymer,$^{1*}$ C. J. McKinstrie,$^{2}$ L. Mejling,$^3$ and K. Rottwitt$^3$}
\address{$^1$Department of Physics, University of Oregon, Eugene, Oregon 97403, USA\\
$^2$Bell Laboratories, Alcatel-Lucent, Holmdel, New Jersey 07733, USA\\ 
$^3$Department of Photonics Engineering, Technical University of Denmark, DK 2800 Kgs. Lyngby, Denmark}
\email{$^*$raymer@uoregon.edu}


\begin{abstract}
We explore theoretically the feasibility of using frequency conversion
by sum- or difference-frequency generation, enabled by
three-wave-mixing, for selectively multiplexing orthogonal input
waveforms that overlap in time and frequency. Such a process would
enable a drop device for use in a transparent optical network using
temporally orthogonal waveforms to encode different channels. We model
the process using coupled-mode equations appropriate for wave mixing
in a uniform second-order nonlinear optical medium pumped by a strong
laser pulse. We find Green functions describing the
process, and employ Schmidt (singular-value) decompositions thereof to
quantify its viability in functioning as a coherent waveform
discriminator. We define a “selectivity” figure of merit in terms of
the Schmidt coefficients, and use it to compare and contrast various
parameter regimes via extensive numerical computations. We identify
the most favorable regime (at least in the case of no pump chirp) and
derive the complete analytical solution for the same. We bound the
maximum achievable selectivity in this parameter space. We show that
including a frequency chirp in the pump does not improve selectivity
in this optimal regime. We also find an operating regime in which
high-efficiency frequency conversion without temporal-shape
selectivity can be achieved while preserving the shapes of a wide
class of input pulses. The results are applicable to both classical
and quantum frequency conversion.
\end{abstract}

%\ocis{(190.0190) Nonlinear optics; (190.4223) Nonlinear wave mixing; (270.0270) Quantum optics.} % REPLACE WITH CORRECT OCIS CODES FOR YOUR ARTICLE

%%%%%%%%%%%%%%%%%%%%%%% References %%%%%%%%%%%%%%%%%%%%%%%%%
\bibliographystyle{osajnl}

%%%%%%%%%%%%%%%%%%%%%%%%%%  body  %%%%%%%%%%%%%%%%%%%%%%%%%%
\section{Introduction}
Efficient multiplexing of signals in and out of multiple optical
channels is central to both quantum and classical optical
communication networks. This is accomplished using an add/drop device
(or filter). The channels are defined by a set of modes, and ideally,
these modes are orthogonal field distributions. For example: a
discrete set of frequencies (wavelength-division multiplexing, WDM),
or time-bins (time-division multiplexing, TDM), or polarization
(polarization-division multiplexing, PDM). True multiplexing, meaning
the ability to efficiently route, add, and drop signals between
channels, can be accomplished using the above-mentioned schemes. A
less powerful form of multiplexing is offered by using schemes that do
not permit efficient signal routing, adding, and dropping, but do
allow detection of signals in different channels. Examples of this are
optical code-division multiple access (OCDMA), quadrature amplitude
modulation (QAM), and optical orthogonal frequency-division
multiplexing (OFDM) \cite{bib01, shieh08}. In those cases the detector
has special capabilities that allow the separate detection of signals
associated with different channels, but cannot efficiently separate
them into distinct spatial channels for routing.

   A goal is to accomplish true orthogonal-waveform multiplexing,
   which would allow signals in different optical channels defined by
   different orthogonal waveforms to be spatially separated
   \cite{ptd12}. The waveforms making up the orthogonal basis set will
   be overlapping in both time and frequency spectra, so standard
   frequency or time separation techniques do not apply. Because
   waveforms, or wave packets, have unique signatures in both time and
   frequency, we will call a system based on such a scheme orthogonal
   time-frequency-division multiplexing (OTFDM).

   Nonlinear optical processes such as three-wave mixing (TWM) have
   previously been applied to quantum networking in the context of
   improving infrared single-photon detection efficiency by
   up-conversion \cite{vand04, albota04, rous04}, and WDM by
   down-conversion \cite{ou10}. An important step toward optical OTFDM
   was made by the group of C. Silberhorn, who proposed an optical
   ‘pulse gate’ based on nonlinear sum-frequency generation by TWM
   \cite{silb11, bib02}. The purpose of the present paper is to
   explore this TWM frequency conversion scheme theoretically, with
   the introduction of analytical tools that help clarify the
   physics. Specifically, we want to check a conjecture that the
   optimum operating regime for shape-selective and efficient
   frequency conversion is that in which one of the signals
   copropagates with the same group velocity as the pump pulse
   \cite{bib02}. The results are applicable to classical and quantum
   frequency conversion. Related results have also been explored for
   the case of four-wave mixing \cite{enk10, hayd10, mej12}.

\section{Equations of motion and figure of merit}

We are concerned with sum/difference frequency generation processes
involving three-wave mixing in any $\chi^{(2)}$-nonlinear medium. We
designate the pulse-carrier frequencies of the three participating
field channels as $\omega_s$, $\omega_r$, and $\omega_p$, where
$\omega_p$ is the strong-pump channel, and $\omega_s$, $\omega_r$ are
the weak signal and idler channels (we assume
$\omega_s<\omega_r$). Though we account for group velocity
mismatch between the channels, we restrict our analysis to
sufficiently narrow-band (broad duration) pulses so as to neglect
higher order effects such as group velocity dispersion.  Starting from
the standard three-wave interaction equations for these channels and
assuming a strong non-depleting pump pulse, an
appropriate choice of channel carriers and polarizations that ensures
energy conservation $(\omega_r = \omega_s+\omega_p)$, and
phase-matching $(k_r-k_s-k_p=0)$ yields the following evolution
equations in the spatio-temporal domain\cite{cerullo03}:
\begin{align*}
(\partial_z+\beta_r\partial_t)A_r(z,t) = i\gamma A_p(t-\beta_pz) A_s(z,t),\tag{1a}\label{1a}\\
(\partial_z+\beta_s\partial_t)A_s(z,t) = i\gamma A_p^*(t-\beta_pz) A_r(z,t),\tag{1b}\label{1b}
\end{align*}
\noindent where for $j\in\{s,r,p\}$, $\beta_j:=\beta^{(1)}(\omega_j)$
are the group slownesses of pulses with carrier-frequency $\omega_j$
(in any arbitrary frame), and $\gamma$ is a measure of the
mode-coupling strength, which is a product of the effective
$\chi^{(2)}$-nonlinearity coefficient and the pump power. We assume
the pump pulse is strong enough that it remains unaltered by the
interaction, but not so strong as to affect the group velocities of
the signal/idler pulses. The mode-amplitudes $A_j(z,t)$ can be
interpreted either as the quantum wavefunction amplitudes in the
single-photon case \cite{smi07}, or as the pulse-envelope functions in
the slow-varying envelope approximation in the classical-pulse limit:
\begin{equation*}
E_p(z,t) = A_p(t-\beta_pz)\exp[i(k_pz-\omega_pt)].\tag{2}\label{2}
\end{equation*}
We assume the field-polarizations of the three channels are fixed for
optimal phase-matching, and hence treat them as scalar fields. The
pump amplitude is square-normalized $\left(\int|A_p(t)|^2dt =
1\right)$. We denote the length of our uniform-medium with $L$, and
assume the interaction starts at $z=0$. The solutions to
Eqs. \eqref{1a} and \eqref{1b} can be represented using the Green function
(GF) formalism, thus:
\begin{equation*}
A_j(L,t)=\sum\limits_{k=r,s}\int\limits_{-\infty}^{\infty}G_{jk}(t,t')A_k(0,t')dt'.\tag{3}\label{3}
\end{equation*}
\noindent where $A_k(0,t')$ are the input amplitudes and $A_j(L,t)$
are the output amplitudes for $j,k \in \{r,s\}$. The overall GF is
unitary, but the block transfer functions ($G_{jk}(...)$) by
themselves, are not. We can affect the GF of the process by varying
the medium length ($L$), pump power ($\gamma$), pump pulse-shape
($A_p(t)$) and the group-slownesses (inverse group velocities) of the
various channels ($\beta_j$).  If the GF is `separable',
i.e. $G_{rs}(t,t')=\Psi(t)\phi^*(t')$, then with sufficient
pump power, an incident {\it s}-channel signal of
temporal shape $\phi(t')$ can be $100\%$ converted into the outgoing
{\it r}-channel packet $\Psi(t)$, and any incoming signal that is
orthogonal to $\phi(t')$ will be left unconverted. In general however,
the GF is not separable. The ability to separate temporal modes
becomes easier to quantify if we represent the GF with its
singular-value decomposition \cite{enk10,hayd10,stra98,gbur}:
\begin{align*}
G_{rr}(t,t') &= \sum\limits_n\tau_n\Psi_{n}(t)\psi_{n}^*(t'),\quad
&G_{rs}(t,t') &= \sum\limits_n\rho_n\Psi_{n}(t)\phi_{n}^*(t'),\tag{4a}\label{4a}\\
G_{ss}(t,t') &= \sum\limits_n\tau_n^*\Phi_{n}(t)\phi_{n}^*(t'),\quad
&G_{sr}(t,t') &= -\sum\limits_n\rho_n^*\Phi_{n}(t)\psi_{n}^*(t').\tag{4b}\label{4b}
\end{align*}
\noindent The functions $\psi_{n}(t'), \phi_{n}(t')$ are the input
``Schmidt modes'' and $\Psi_{n}(t), \Phi_{n}(t)$ are the corresponding
output Schmidt modes for the {\it r} and {\it s} channels
respectively. These functions are uniquely determined by the GF, and
form orthonormal bases in their relevant channels. The
``transmission'' and ``conversion'' Schmidt-coefficients (singular
values) $\{\rho_n\}$ and $\{\tau_n\}$ are constrained by
$|\tau_n|^2+|\rho_n|^2=1$ to preserve unitarity. It is convenient to
choose the mode-index `n' in decreasing order of Schmidt mode
conversion-efficiency (CE) ($|\rho_n|^2$). The process is deemed
perfectly mode-selective if the CE $|\rho_1|^2=1$, and $|\rho_m|=0;$
for every $ m\neq 1$ (i.e., the TWM process performs full frequency
conversion on one particular input mode and transmits all power from
any orthogonal mode in the same input channel). We can quantify the
add/drop quality of the GF using the ordered-set of conversion
efficiencies to define an add/drop `selectivity':
\begin{equation*}
S :=\frac{|\rho_1|^4}{\sum\limits_{n=1}^\infty|\rho_n|^2}\leq 1.\tag{5}\label{5}
\end{equation*}
We call the factor $|\rho_1|^2/(\sum_{n=1}^\infty|\rho_n|^2)$ the
`separability', and the additional multiplier ($|\rho_1|^2$) is the CE
of the dominant temporal mode. The selectivity characterizes both the
degree of separability of the GF and the process efficiency.  The
equality in Eq. \eqref{5} holds for a perfect add/drop device.
The unitary nature of the transformation imposes a pairing between the
Schmidt modes across the {\it r} and {\it s} channels
\cite{enk10,braun05}. Consider arbitrary input and output fields
expressed as discrete sums over corresponding Schmidt
modes:
\begin{align*}
A_r(t)|_{\text{in}}=\sum\limits_n a_{n}\psi_{n}(t),\quad
A_r(t)|_{\text{out}}=\sum\limits_n c_{n}\Psi_{n}(t),\tag{6a}\label{6a}\\
A_s(t)|_{\text{in}}=\sum\limits_n b_{n}\phi_{n}(t),\quad
A_s(t)|_{\text{out}}=\sum\limits_n d_{n}\Phi_{n}(t).\tag{6b}\label{6b}
\end{align*}
\noindent The coefficients $\{a_n, b_n, c_n, d_n\}$ are pairwise
related via a unitary beam-splitter-like transformation \cite{enk10},
which, if we assume real $\tau_n$ and $\rho_n$, are expressed as:
\begin{align*}
c_{n}&= \tau_na_{n}+\rho_nb_{n},\tag{7a}\label{7a}\\
d_{n}&= \tau_nb_{n}-\rho_na_{n},\tag{7b}\label{7b}
\end{align*}
\noindent where the $n^{\text{th}}$-Schmidt mode CE ($|\rho_n|^2 =
1-|\tau_n|^2$) is analogous to ``reflectance''. All time-domain
functions described thus far have corresponding frequency-domain
analogs. The form of the GF in frequency domain can also provide
meaningful insights. If we define functions
$\widetilde\Psi_{n}(\omega)$ and $\widetilde\phi_{n}(\omega)$ as the
Fourier-transforms of the corresponding time-domain Schmidt modes
$\Psi_{n}(t)$ and $\phi_{n}(t)$, then:
\begin{equation*}
\widetilde{G}_{rs}(\omega,\omega')=\int dt\int dt'\exp[i\omega
  t]G_{rs}(t,t')\exp[-i\omega't']=\sum\limits_n\rho_n\widetilde\Psi_{n}(\omega)\widetilde\phi^*_{n}(\omega').\tag{8}\label{8}
\end{equation*}
\noindent The above analysis has been shown \cite{enk10, hayd10} to
apply equally well to quantum wave-packet states as to classical
fields, for the simple reason that all the relations are linear in the
mode creation and annihilation operators. Thus the GFs found here can
model experiments on frequency conversion (FC) of single-photon
wave-packet states \cite{ptd12, rakher10, ikuta11} or FC of other
quantum states such as squeezed states containing multiple photons.

\section{Low-conversion limit}
We can develop an important guide to the different regimes of TWM by
solving the problem for small interaction strengths ($\gamma$) for
arbitrary group slownesses and pulse shapes (following the discussion
in \cite{mej12}). We define the coupling coefficient as
$\kappa(z,t)=\gamma A_p(t-\beta_pz)$. For this calculation, we could
allow the nonlinearity $\gamma(z)$ to be position dependent, which can
be used as a design feature if desired \cite{silb11,bran11}, but for
simplicity we continue to assume that the medium is uniform. By
integrating Eqs. \eqref{1a} and \eqref{1b} with respect to $z$, we get the
exact relations:
\begin{align*}
A_r(L,t)=A_r(0,t-\beta_rL)+i\int\limits_0^Ldz'\kappa(z',t)A_s(z',t_r'),\tag{9a}\label{9a}\\ 
A_s(L,t)=A_s(0,t-\beta_sL)+i\int\limits_0^Ldz'\kappa^*(z',t)A_r(z',t_s'),\tag{9b}\label{9b}
\end{align*}
where $t_r':=t-\beta_r(L-z')$ and $t_s':=t-\beta_s(L-z')$. Treating
the coupling as a perturbation, we get
\begin{align*}
A_r(L,t)\approx A_r(0,t_r)+i\int\limits_0^Ldz'\kappa(z',t)A_s(0,t_r),\tag{10a}\label{10a}\\
A_s(L,t)\approx A_s(0,t_s)+i\int\limits_0^Ldz'\kappa^*(z',t)A_r(0,t_s),\tag{10b}\label{10b}
\end{align*}
where $t_r=t-\beta_rL$, $t_s=t-\beta_sL$. The $\approx$ symbols
indicate that perturbative approximations render Eqs. \eqref{10a} and
\eqref{10b} weakly non-unitary. By defining
$t'=t-\beta_rL+\beta_{rs}z'$, where $\beta_{rs}=\beta_r-\beta_s$ is
the difference in slownesses, one can change the integration variable
to time, and rewrite Eqs. \eqref{10a} and \eqref{10b} using the
approximate Green function $\overline{G}_{jk}(t,t')$ in the
low-conversion limit \cite{mej12}:
\begin{equation*}
A_j(L,t)\approx A_j(0,t_j)+\int\limits_{-\infty}^\infty dt'\overline{G}_{jk}(t,t')A_k(0,t')\bigg\bracevert_{k\neq j},\tag{11}\label{11}
\end{equation*}
\begin{align*}
\overline{G}_{rs}(t,t')&=i\frac{\gamma}{\beta_{rs}}A_p\left(\frac{\beta_{rp}t'-\beta_{sp}(t-\beta_rL)}{\beta_{rs}}\right)H(t'-t+\beta_rL)H(t-t'-\beta_sL),\tag{12a}\label{12a}\\
\overline{G}_{sr}(t,t')&=-i\frac{\gamma^*}{\beta_{rs}}A_p^*\left(\frac{\beta_{rp}(t-\beta_sL)-\beta_{sp}t'}{\beta_{rs}}\right)H(t'-t+\beta_sL)H(t-t'-\beta_rL),\tag{12b}\label{12b}
\end{align*}
\noindent where $t_j=t-\beta_jL$, $\beta_{jp}=\beta_j-\beta_p;\forall
j\in\{r,s\}$, and $H(x)$ is the Heaviside step-function.

\begin{figure}[htb]
\centering
\includegraphics[width=3.2in]{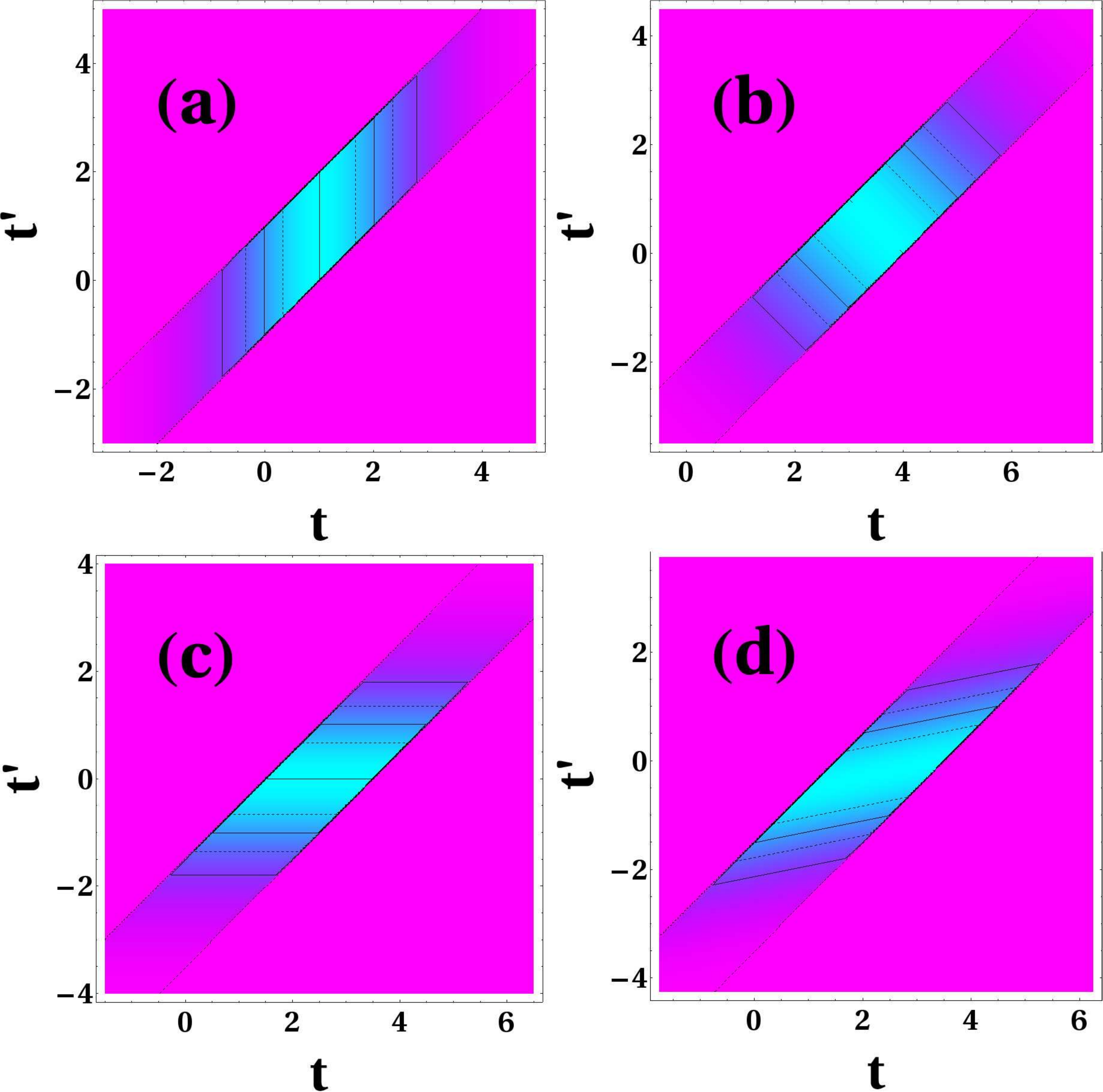}
\caption{Green function $\overline{G}_{rs}(t,t')$ in low-conversion
  limit for medium length $L=1$ and Gaussian pump duration
  $\tau_p=1$. {\bf(a)} $\beta_r=\beta_p=1,$ $\beta_s=-1,$ {\bf(b)}
  $\beta_r=4,$ $\beta_s=2,$ $\beta_p=3,$ {\bf(c)} $\beta_r=3.5,$
  $\beta_s=\beta_p=1.5,$ {\bf(d)} $\beta_r=3.5,$ $\beta_s=1.5,$
  $\beta_p=1$}
\label{fig01}
\end{figure}

Eqs. \eqref{12a} and \eqref{12b} provide a simple way to
understand the FC process for arbitrary relations between group
slownesses. A first observation is that the Heaviside step-functions
represent that because the medium length is finite, the time of
interaction is restricted to the {\it $t'$} interval
$t'\in(t-\beta_rL,t-\beta_sL)$. In the ({\it $t,t'$}) domain, this
interval corresponds to a $45^\circ$-tilted band with width
$\beta_{rs}L$. Therefore, if the goal is to have the GF separable in
{\it $t$} and {\it $t'$}, the shape of this interval poses a
challenge. Whatever the pulse shape of the pump, the low-conversion
Green function is proportional to a scaled version of that shape. Note
that if ($\beta_{rp}=0$ or $\beta_{sp}=0$), then the factor $A_p(...)$
in Eqs. \eqref{12a} and \eqref{12b} depends only on {\it $t$} (or {\it
  $t'$}), making that factor in the GF separable. Further insight is
obtained by plotting the GF, as in Fig. \ref{fig01}, for the case of a
(normalized) Gaussian pump pulse
$A_p(t)=\left(\tau_p^2\pi\right)^{-1/4}\exp[(-t^2/(2\tau_p^2)]$ with
duration $\tau_p$.

For the four GF's plotted in Fig. \ref{fig01}, the computed CE's for
the first four temporal modes are listed in Table \ref{tbl1}, where
$\overline\gamma = \gamma/\beta_{rs}$ is of order $0.01$. The
corresponding selectivities are: $S =$ {\bf(a)}
$0.646\overline\gamma^2$, {\bf(b)} $0.676\overline\gamma^2$, {\bf(c)}
$0.646\overline\gamma^2$, {\bf(d)} $0.610\overline\gamma^2$. The
Schmidt coefficients were numerically computed by performing a
singular value decomposition (SVD) of the GF in Eq. \eqref{12a}.

The slope of the line along the highest part of the band (lightest
color) is given by
\begin{equation*}
\text{slope}=\frac{dt'}{dt}\bigg\bracevert_{\text{max}}=\frac{\beta_s-\beta_p}{\beta_r-\beta_p}.\tag{13}\label{13}
\end{equation*}
\begin{table}[htb]
\centering
\caption{Conversion efficiencies for the first four dominant Schmidt
  modes for the Green functions from Fig.
  \ref{fig01}. $\overline\gamma = \gamma/\beta_{rs}$.}
\begin{tabular}{|c|c|c|c|c|}
\hline
{\bf (a)} & $1.0\overline\gamma^2$ &  $0.306\overline\gamma^2$ & $0.088\overline\gamma^2$ & $0.037\overline\gamma^2$\\
\hline
{\bf (b)} & $1.0\overline\gamma^2$ & $0.275\overline\gamma^2$ & $0.064\overline\gamma^2$ & $0.033\overline\gamma^2$\\
\hline
{\bf (c)} & $1.0\overline\gamma^2$ & $0.306\overline\gamma^2$ & $0.088\overline\gamma^2$ & $0.037\overline\gamma^2$\\
\hline
{\bf (d)} & $1.0\overline\gamma^2$ & $0.342\overline\gamma^2$ & $0.115\overline\gamma^2$ & $0.047\overline\gamma^2$\\
\hline
\end{tabular}
\label{tbl1}
\end{table} 

\begin{figure}[htb]
\centering
\includegraphics[width=3.2in]{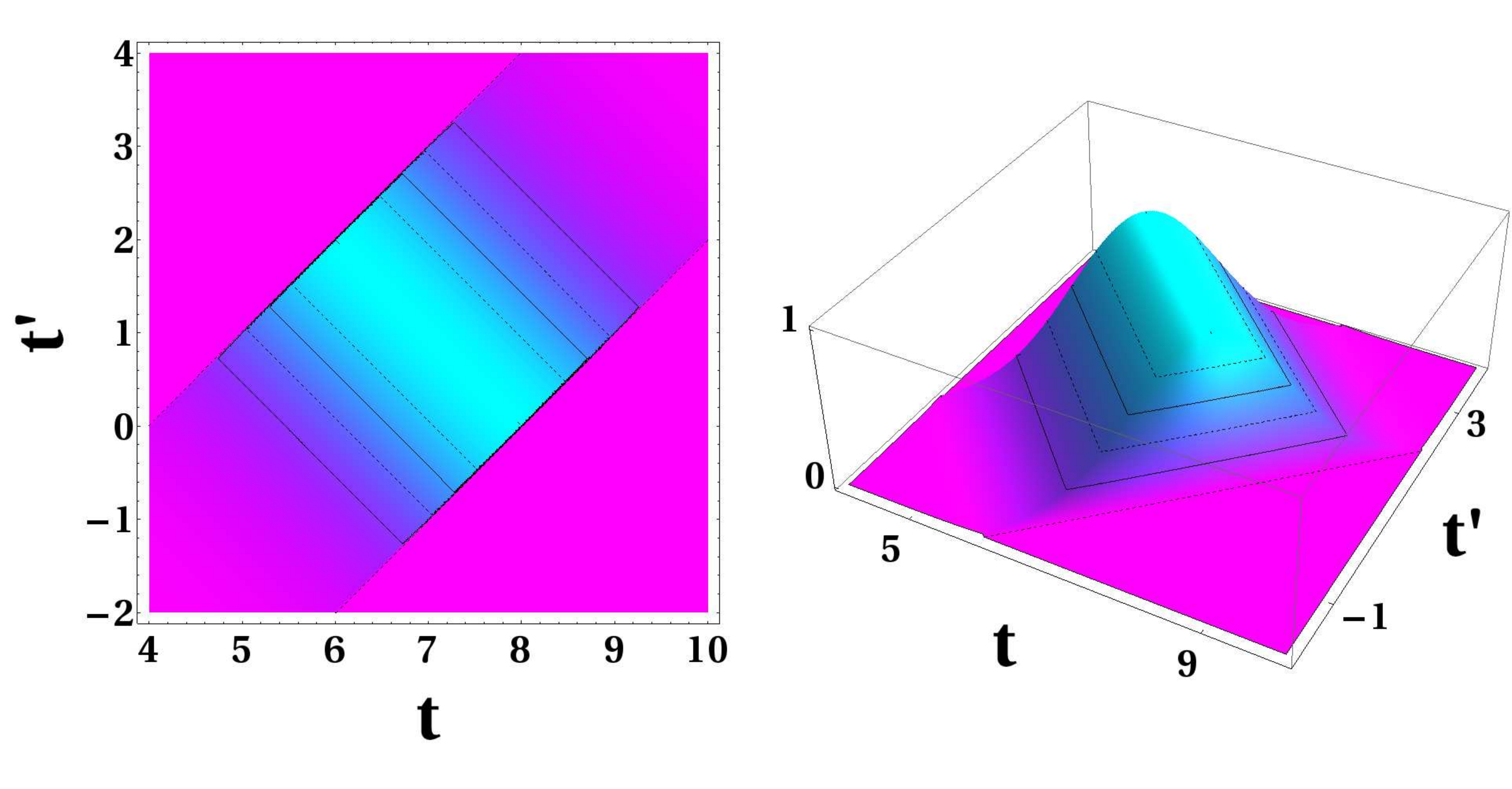}
\caption{Low-conversion Green function for $\beta_r=8,$ $\beta_s=4,$
  $\beta_p=6,$ $\tau_p=0.707,$ $L=1$. Top view and perspective view.}
\label{fig02}
\end{figure}

\begin{figure}[htb]
\centering
\includegraphics[width=3.2in]{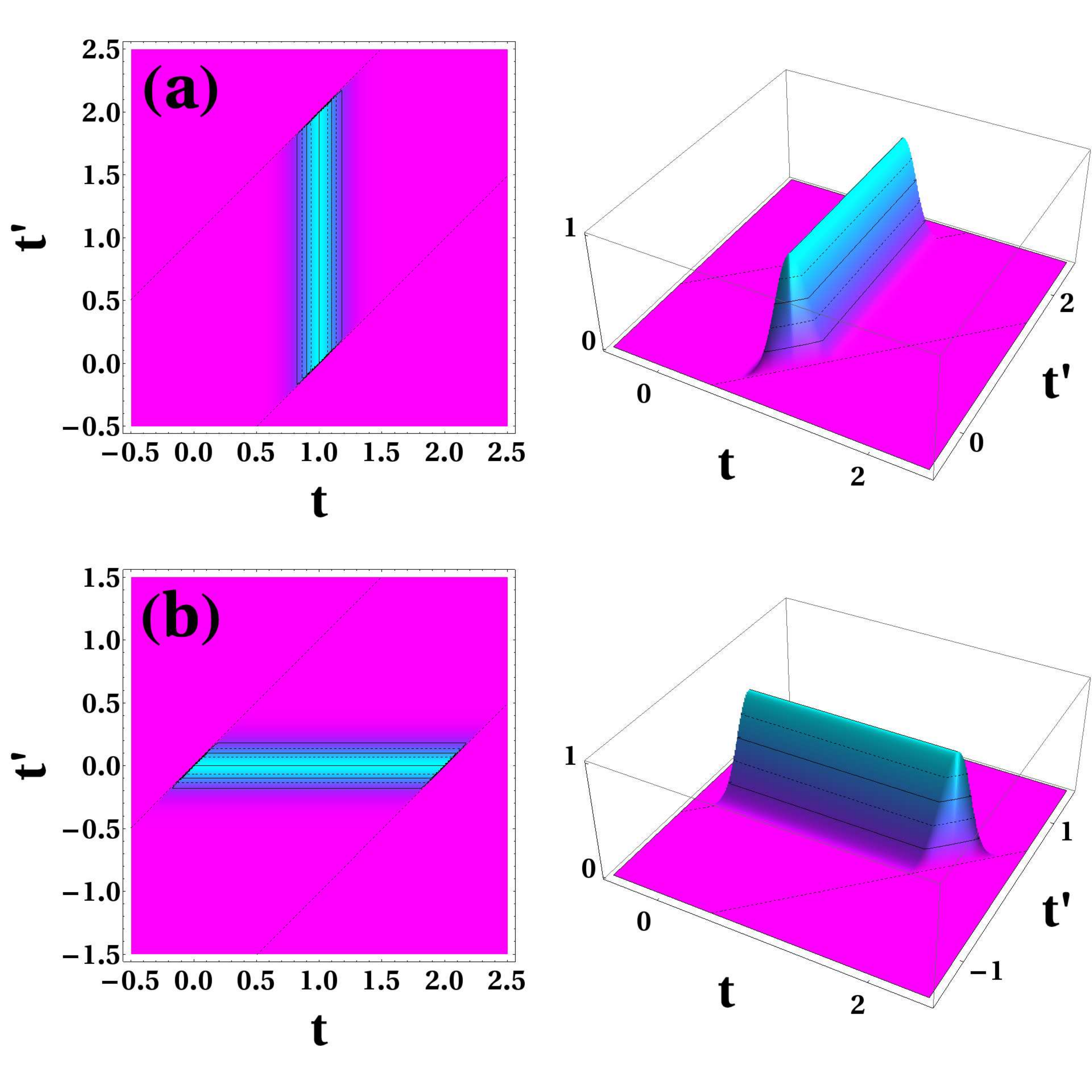}
\caption{Low converison Green function for {\bf(a)}
  $\beta_r=\beta_p=1,$ $\beta_s=-1,$ $\tau_p=0.1,$ $L=1$. For {\bf(b)}
  $\beta_r=2$, $\beta_s=\beta_p=0,$ $\tau_p=0.1,$ $L=1$.}
\label{fig03}
\end{figure}

In an attempt to create an approximately separable GF, one can choose
parameters as in Fig. \ref{fig02}. The computed CE's for the first
four temporal modes in this case are $\{1.0\overline\gamma^2,
0.029\overline\gamma^2, 0.028\overline\gamma^2,
0.011\overline\gamma^2\}$, where $\overline\gamma = \gamma/\beta_{rs}$
is of order $0.01$. The selectivity is $S =
0.913\overline\gamma^2$. While the separability is high, the CE of the
first Schmidt mode is of the order of $\overline\gamma^2$.
Improved selectivity can be achieved using the strategy proposed in
\cite{bib02}, where one of the signals is matched in slowness to the
pump, as in Fig. \ref{fig01}(a), and the pump pulse is made very
short. The short pump width helps counter the ill effects of the
$45^\circ$-sloping step-functions on GF separability by selecting a
narrow vertical or horizontal region in ($t, t'$) space. These choices
give the GF's (in the low-CE limit) in Fig. \ref{fig03}.
The numerically computed CE's for the first four temporal modes in
Fig. \ref{fig03}(a) are $\{1.0\overline\gamma^2,
0.022\overline\gamma^2, 0.006\overline\gamma^2,
0.003\overline\gamma^2\}$, and the selectivity is $S =
0.967\overline\gamma^2$. In Fig. \ref{fig03}(b) the CE's and the
selectivities are identical to case \ref{fig03}(a). $\overline\gamma$
is of order $0.01$.

\begin{figure}[htb]
\centering
\includegraphics[width=3.2in]{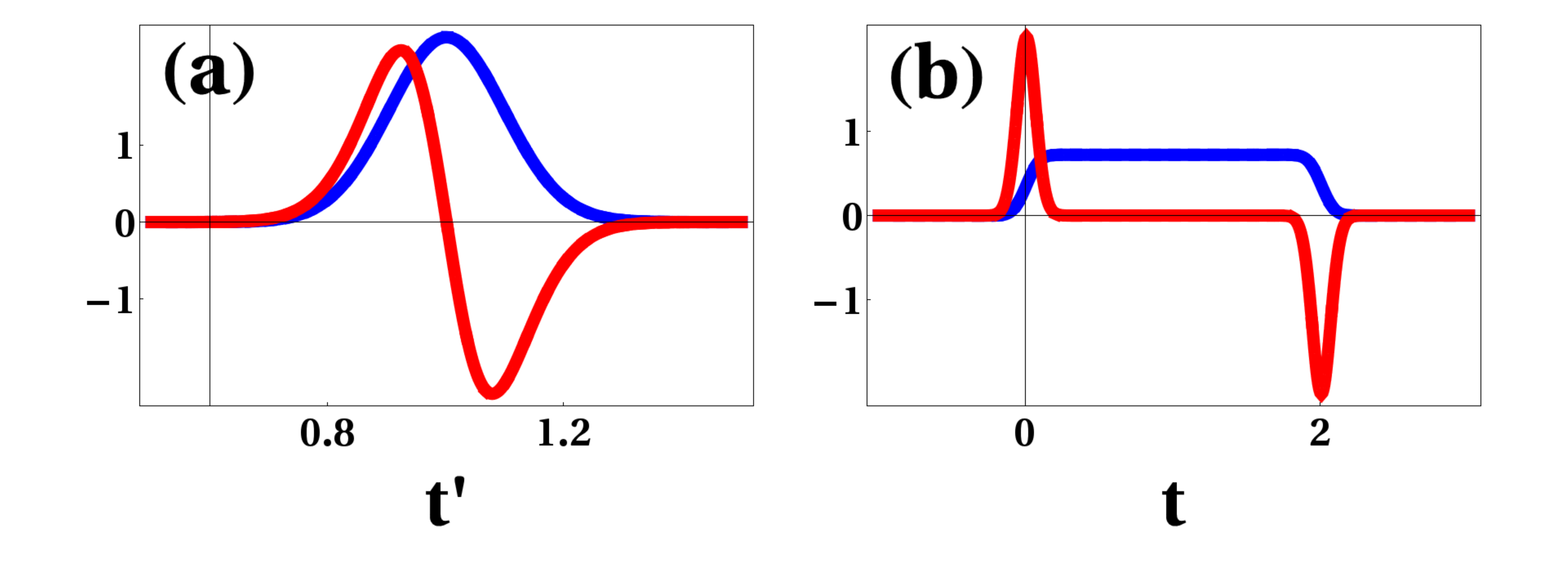}
\caption{Temporal Schmidt modes for the case in Fig. \ref{fig03}(b)
  $\beta_r=\beta_p=1,$ $\beta_s=-1,$ $\tau_p=0.1,$ $L=1$. {\bf(a)}
  First two input modes. {\bf(b)} Corresponding output modes. First
  modes are in blue. Second modes are in red.}
\label{fig04}
\end{figure}

The temporal Schmidt modes for the case in Fig. \ref{fig03}(b) are
shown in Fig. \ref{fig04}. It is seen that the input modes mimic
the projection of the GF onto the $t'$-axis, while the output modes mimic
the projection onto the $t$-axis. If the input field occupies only the
dominant ($j=1$) Gaussian-like mode, then it is frequency converted with
efficiency $|\rho_1|^2\approx\overline\gamma^2$ and
generates an output pulse that is much longer and rectangular in
shape. Such pulse shaping may or may not be desired, depending on the
application.

The {\it s}-output modes for case \ref{fig03}(a) and the {\it r}-output modes
for case \ref{fig03}(b) will have temporal width $\beta_{rs}L$, which
is the maximum duration of interaction within the medium. Since the
pump copropagates with a matched slowness with one of the input
channels, and the CE is low enough to prevent input-channel depletion,
FC occurs throughout the traversed medium length, stretching the
generated output mode in the other channel due to difference in
slownesses ($\beta_{rs}$).

\begin{figure}[htb]
\centering
\includegraphics[width=3.2in]{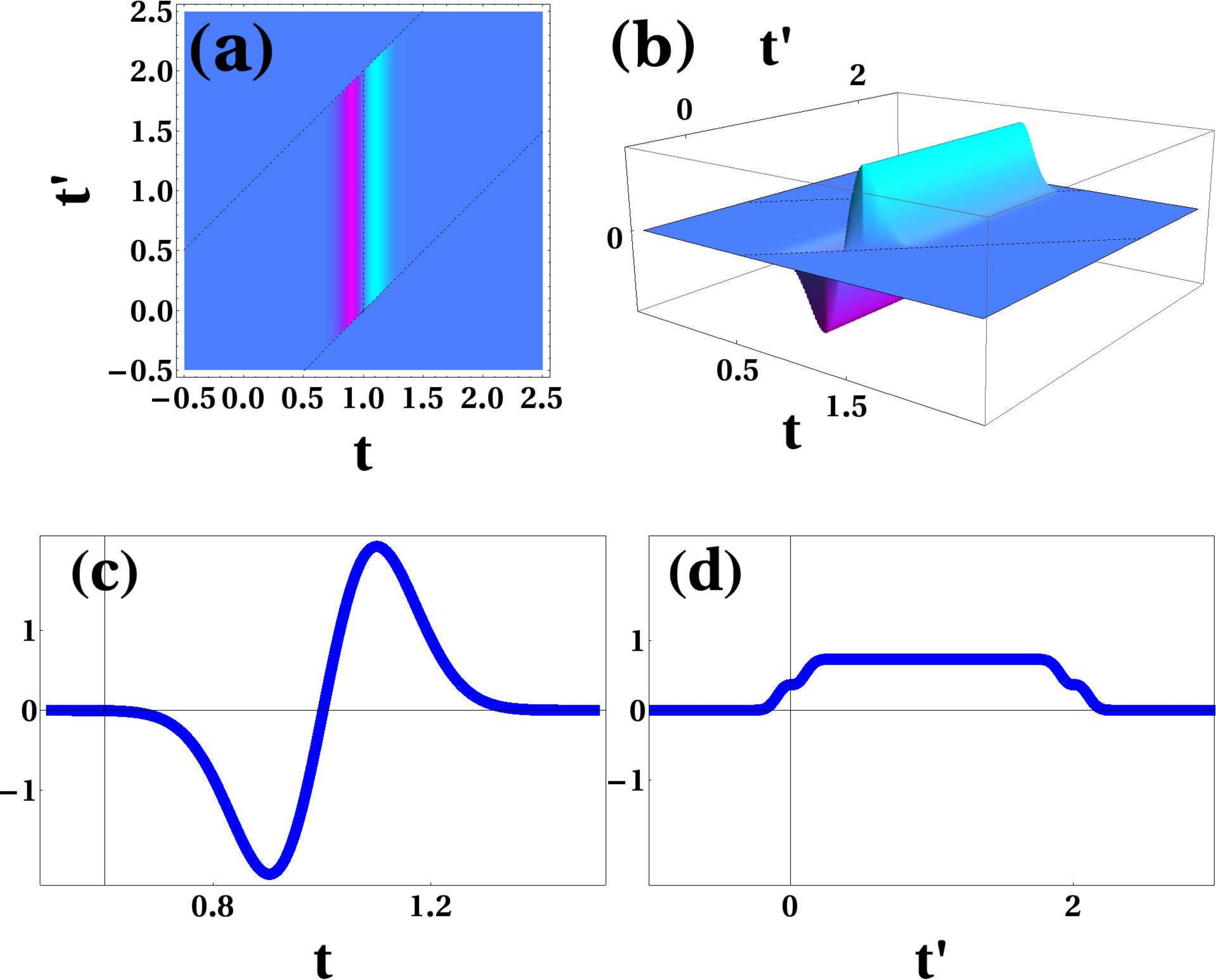}
\caption{{\bf(a), (b)} Low-conversion Green function with higher-order
  pump pulse and $\beta_r=\beta_p=1$, $\beta_s=-1$,
  $\tau_p=0.1$, $L=1$. {\bf(c)} Dominant input Schmidt mode. {\bf(d)}
  Dominant output Schmidt mode.}
\label{fig05}
\end{figure}

To demonstrate the ability to choose which temporal mode is selected
for FC, Fig. \ref{fig05} shows the results for a pump pulse with the
shape proportional to a first-order Hermite-Gaussian function
$HG_1(x)\propto x\exp[-x^2/2]$, which has a zero-crossing at its
``midpoint''. The efficiencies are $\{1.0\overline\gamma^2,
0.049\overline\gamma^2, 0.007\overline\gamma^2,
0.005\overline\gamma^2\}$ and the selectivity is $S =
0.936\overline\gamma^2$, where $\overline\gamma = \gamma/\beta_{rs}$
is of order $0.01$. The dominant mode has a shape similar to the pump
pulse.

Alternatively, the preceding analysis can be carried out in the
frequency-domain, where the GF takes the form:
\begin{align*}
\widetilde{\overline{G}}_{rs}(\omega,\omega')
&=i\frac{\gamma}{\beta_{rs}}\widetilde{A}_p(0,\omega-\omega')\exp[-iL\beta_r(\omega-\omega')\beta_{sp}/\beta_{rs}]\times
\frac{\sin(\overline{\omega}'\beta_{rs}L)}{\overline{\omega}'}\exp[iL\overline{\omega}'(\beta_r+\beta_s)]\\
&=g_1(\omega-\omega')\times g_2(\overline{\omega}'),\tag{14}\label{14}
\end{align*}
where $\overline{\omega}' =
(\beta_{rp}\omega-\beta_{sp}\omega')/(2\beta_{rs})$. Varying the pump
duration $\tau_p$ changes the bandwidth of factor
$g_1(\omega-\omega')$, and the choice of slowness ($\beta_j$) and
medium length ($L$) affects the slope and the phase-matching bandwidth
of factor $g_2(\overline\omega')$ in ($\omega,\omega'$) space. The
separability of $\overline{G}_{rs}(t,t')$ is also evident in
$(\omega,\omega')$-space. As pointed out in \cite{enk10}, for the case
in Fig. \ref{fig03}(a) with $\beta_{rp}=0$, $g_2(\overline\omega')$
would be a sinc-function parallel to the $\omega$-axis with a
phase-matching bandwidth proportional to $1/(\beta_{sp}L)$ [a measure
  of the vertical separation between the edges of the heaviside-step
  functions in Fig. \ref{fig03}(a)], and the shortness of the pump
will cause $g_1(\omega-\omega')$ to have a wide-bandwidth,
intersecting $g_2(\overline\omega')$ at a $45^\circ$
inclination. Alternatively, one could choose parameters such that
$\beta_{sp}=-\beta_{rp}$, giving $g_2(\overline\omega')$ a $-45^\circ$
inclination. In the frequency domain, if the pump bandwidth and medium
length are optimized, we can reproduce a roughly separable GF as in
Fig. \ref{fig02}, with $(t, t')$ replaced by $(\omega,\omega')$. GF
separability suffers in this regime if the pump bandwidth is made much
larger than the phase-matching bandwidth.

To summarize, in the low CE limit, we are able to achieve high
temporal mode separability when the GF is nearly separable
(Fig. \ref{fig03}) and moderate separability when the GF has a
modicum of symmetry (Fig. \ref{fig02}). But the low CE diminishes
the selectivity. We next address whether high selectivity can also
be found in cases with higher conversion efficiencies.
\section{High-conversion regimes}
Assuming energy conservation and perfect phase-matching for the
channel carrier frequencies, the choice of waveguide/material
dispersion is reflected in our equations via the relative magnitudes
of the channel group slownesses. We classify the different
regimes of operation as follows:
\begin{itemize}
\item {\bf Single sideband velocity matched}\\
SSVM: $\beta_s=\beta_p\neq\beta_r$ or $\beta_r=\beta_p\neq\beta_s$
\item {\bf Symmetrically counter-propagating}\\
SCuP: $\beta_{rp}=-\beta_{sp}$
\item {\bf Counter-propagating signals}\\
CuP: $\beta_{sp}\beta_{rp}<0, \quad\beta_s\neq\beta_r$
\item {\bf Co-propagating signals}\\
CoP: $\beta_{sp}\beta_{rp}>0,\quad\beta_s\neq\beta_r$
\item {\bf Exactly co-propagating}\\
ECoP: $\beta_r=\beta_s$
\end{itemize}
In this section, we employ numerical techniques similar to those used
in \cite{hayd10} to construct the GF for any given set of pump
parameters. To accomplish this we numerically propagate a large number
of `test signals' through the medium (chosen to be members of a
complete, orthonormal set of Hermite-Gaussian functions of appropriate
temporal width) to find the effects of the process on an arbitrary
input. This method (described in Appendix I) enables a
comprehensive study of TWM, even for cases for which analytical
solutions are not known.

 We first present numerical results for the SSVM regime, which has
 been favored by C. Silberhorn's group \cite{silb11}, and has yielded
 the best results in terms of selectivity.
\subsection{($\beta_{sp}=0, \beta_{rs}\neq 0$) Single sideband velocity matched regime}
The function of an effective add/drop device is to efficiently
discriminate between orthogonal temporal modes. Since any
channel input enters and traverses through the waveguide in causal
sequence (linearly with the pulse function argument), to achieve
discrimination the pump pulse must overlap with all segments of the
input pulse for a non-zero amount of time within the waveguide. This
ensures that: (a) all the power distributed among all the segments of
the first input Schmidt mode has a chance of interacting with the pump
and being FC'd into the other channel, and (b) the device ``measures''
the entire shape of the temporal input mode, which is essential for
discriminating between different temporal mode shapes.
\begin{figure}[htb]
\centering
\includegraphics[width=3.8in]{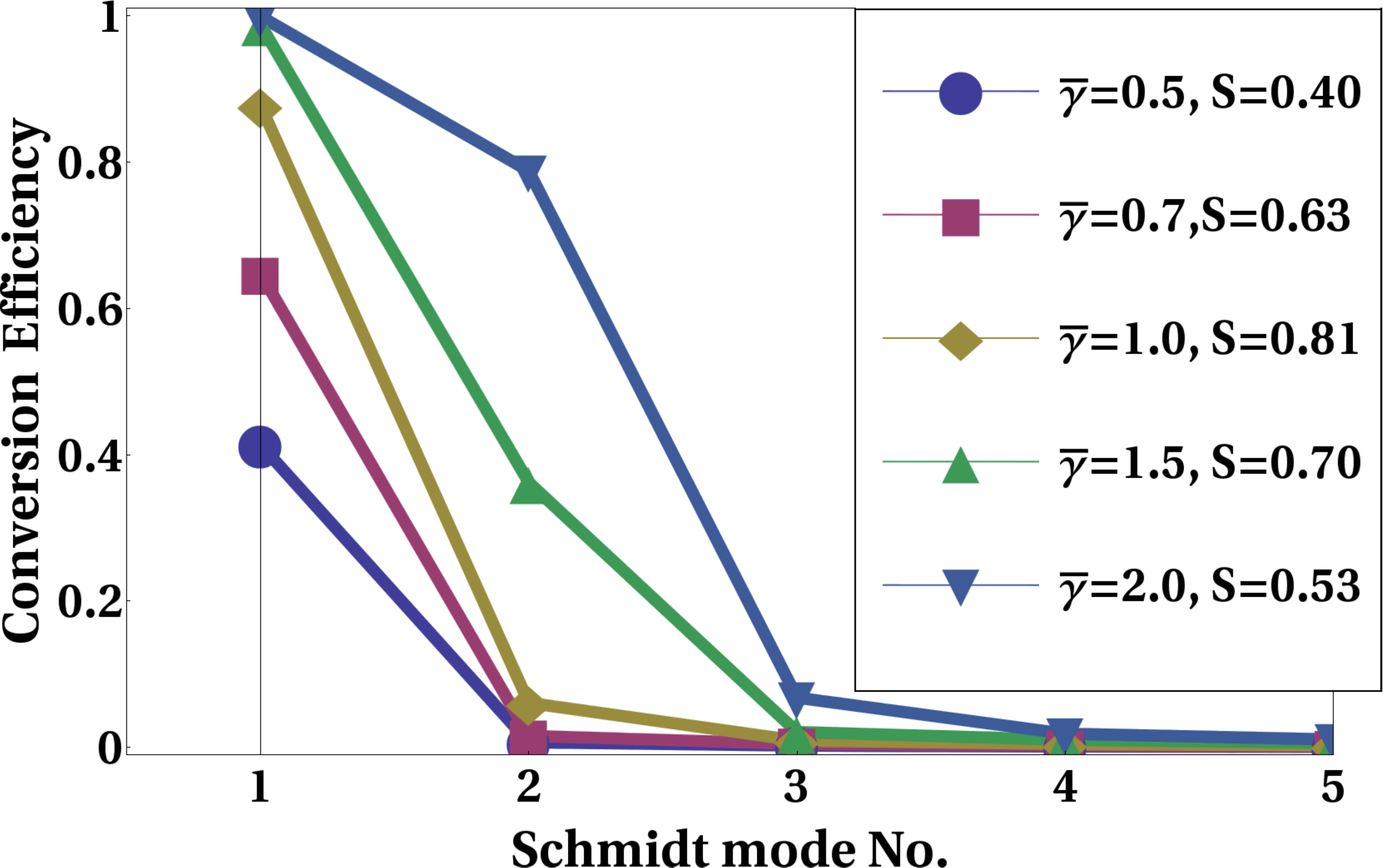}
\caption{Numerically determined conversion efficiencies of the first
  five Schmidt modes for the SSVM case in Fig. \ref{fig03}(b) for
  various $\overline\gamma$. The resulting selectivities {\it S} is given
  in the legend.}
\label{fig06}
\end{figure}
\begin{figure}[htb]
\centering
\includegraphics[width=3.7in]{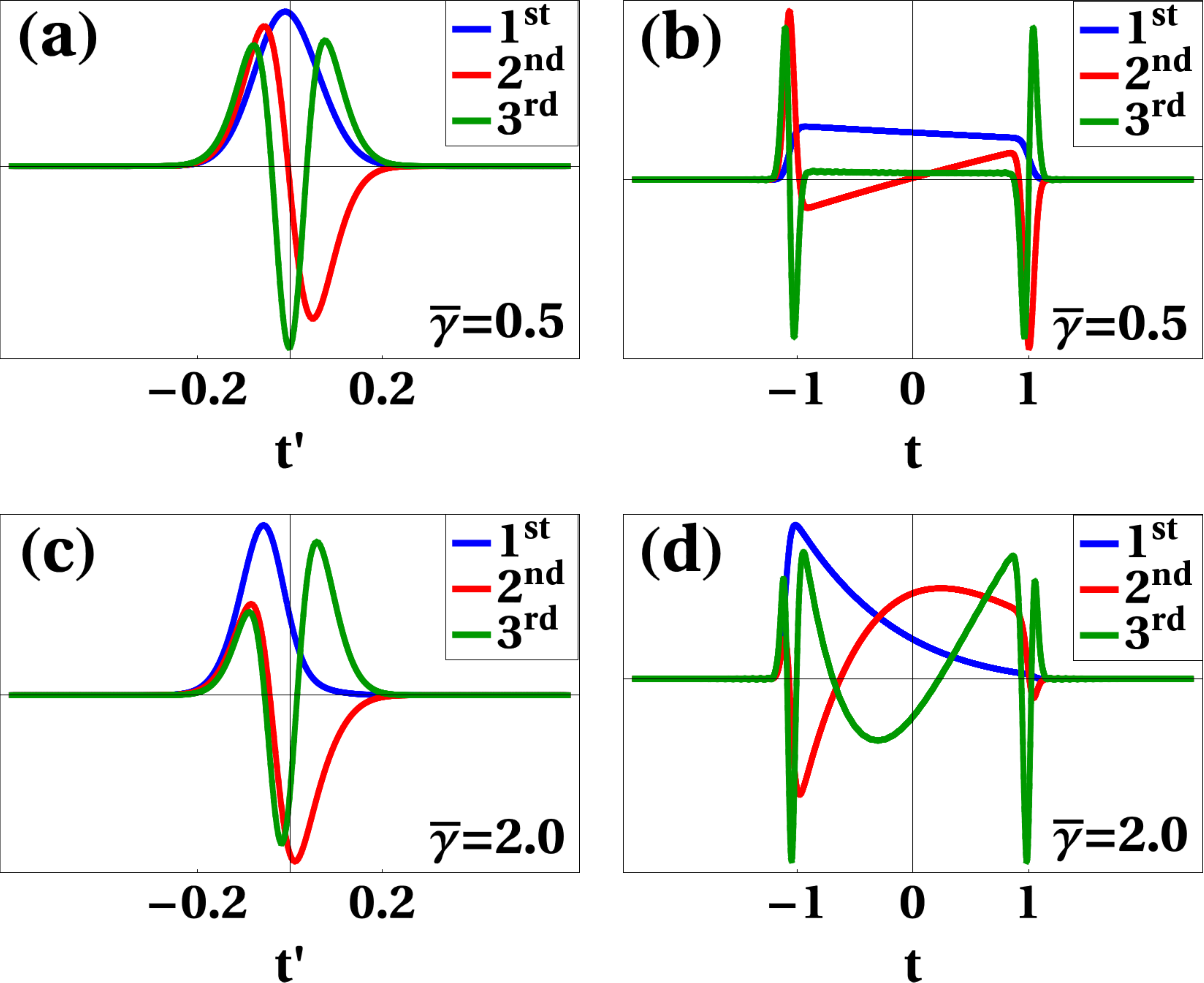}
\caption{The first three {\it s} input (a, c) and {\it r} output (b, d)
  Schmidt modes for $\overline\gamma =$ 0.5(a, b), 2.0(c, d), for
  parameters from Fig. \ref{fig03}(b). Numerical results.}
\label{fig07}
\end{figure}

Two orthogonal temporal modes (say in channel {\it r}) can have
locally similar shapes in certain segments. When these segments
overlap with the pump pulse within the nonlinear medium, the only way
for the device to react to them differently is for the local
instantaneous mode features in channel {\it s} to differ (Eq.
\eqref{1a}), which is determined by all the TWM that has occurred
until that time instant. Both of these intuitive requirements are
satisfied if one of the channel slownesses is matched to the pump
slowness (single sideband group-velocity matched or SSVM), and the
temporal pump width is much shorter than the interaction time
$\beta_{rs}L$. The preceding low CE-limit analysis has already deemed
this regime favorable for separability, and other groups \cite{bib02}
have predicted significant success at higher CE's as well. We now
present the numerical results for the same. We present the complete
exact-analytical solution for the SSVM case in section 5. In the SSVM
regime, for a given pump shape, the selectivity is influenced most by
the GF aspect ratio ($\tau_p/(\beta_{rs}L)$) and effective interaction
strength ($\overline\gamma=\gamma/\beta_{rs}$).

In Fig. \ref{fig06}, we plot the numerically determined CE for the
first five Schmidt modes for various $\overline\gamma$ for a Gaussian
pump-pulse with parameters from Fig. \ref{fig03}(b) ($\beta_r=2$,
$\beta_s=\beta_p=0$, $\tau_p=0.1$, $L=1$). The selectivity values are
listed in the inset. A maximum selectivity of $0.81$ is found for
$\overline\gamma=1.0$.

\begin{figure}[htb]
\centering
\includegraphics[width=3.2in]{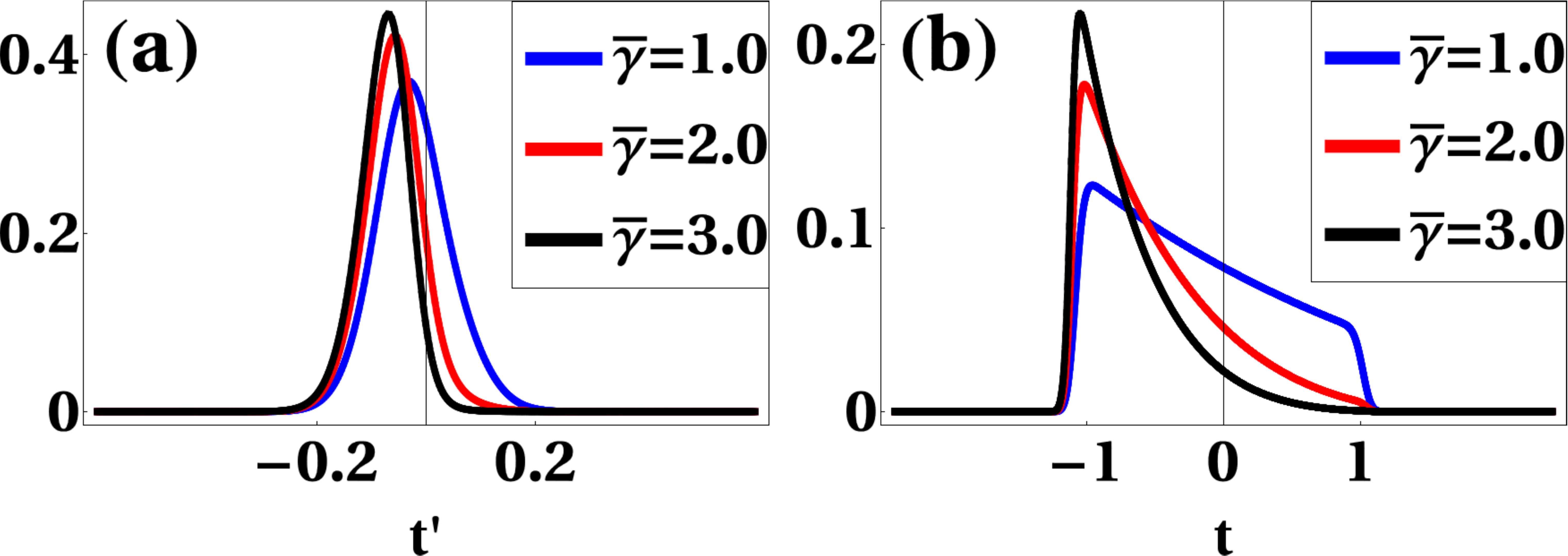}
\caption{Distortion of the first Schmidt modes ({\it r} input (a) and
  {\it s} output (b)) with increasing $ \overline\gamma$, for parameters
  from Fig. \ref{fig03}(b). Numerical results.}
\label{fig08}
\end{figure}

Although the GF displays good mode-separability at low-CE's, the
selectivity is unable to maintain high values beyond a certain
$\overline\gamma$. Figure \ref{fig07} shows the first three input and
output Schmidt modes for $G_{rs}(t,t')$ for the same
case. Figure \ref{fig08} shows the first Schmidt modes from
both channels for increasing $\overline\gamma$. Note the strong shape
distortion relative to the low-CE case, reflecting the
change in the GF shape with increasing $\overline\gamma$. This
illustrates the limits of validity of the approximation used in
\cite{bib02}.

Shortening the pump width by a factor of $10$ minutely improves the
selectivity, while lengthening pump width causes it to decrease. We
present the analytical solution for this SSVM regime in section 5,
where we show that this case leads to the highest selectivity of all
the regimes treated in this study.

\subsection{($\beta_{rp} = -\beta_{sp}$) Symmetrically counter-propagating signals regime, shape preserving FC}

\begin{figure}[htb]
\centering
\includegraphics[width=3.2in]{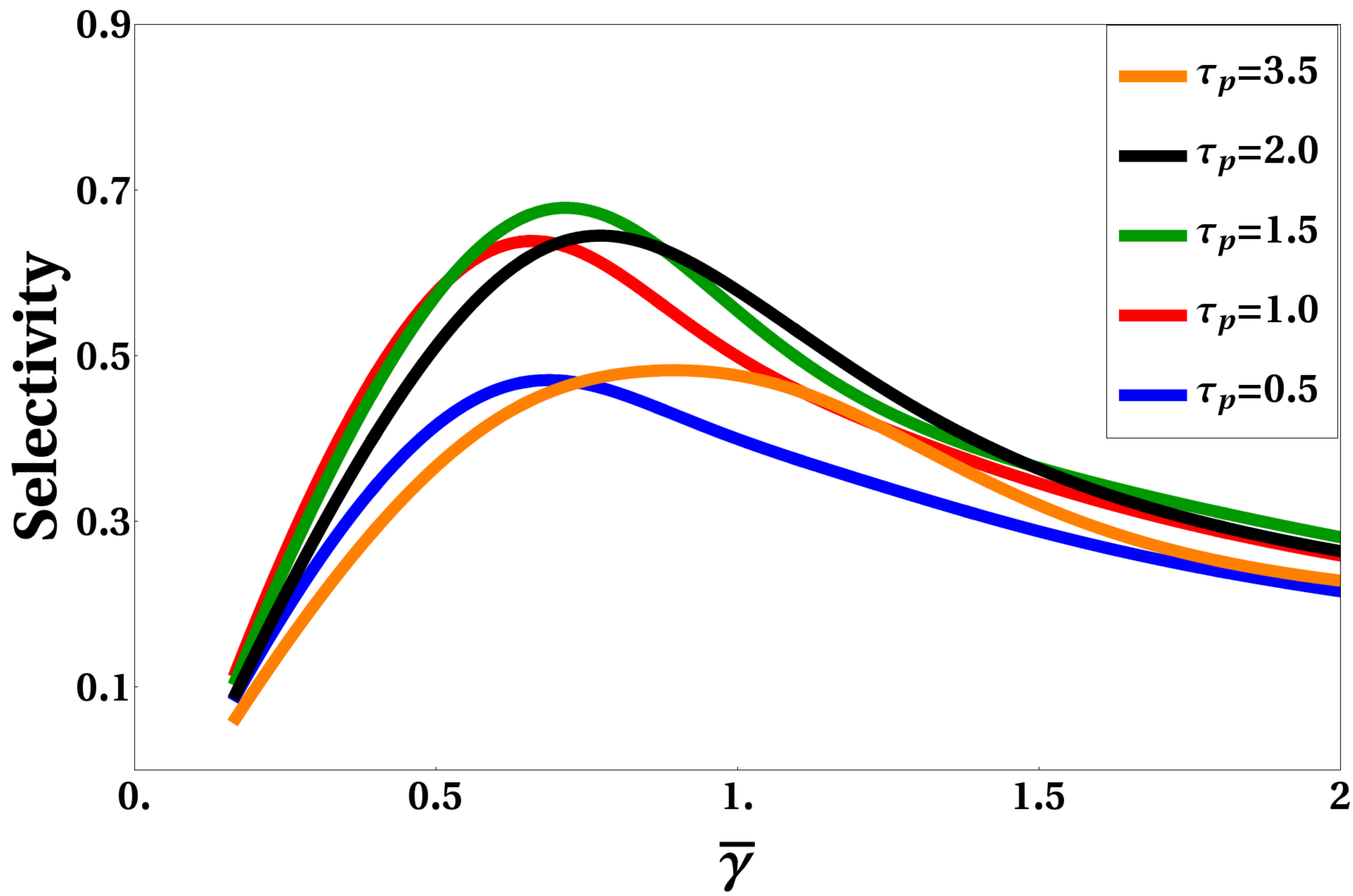}
\caption{Selectivity vs. $\overline\gamma$ for Gaussian pumps of
  various widths. $\beta_s = 0$, $\beta_p = 2$, $\beta_r = 4$, $L =
  1$. Numerical results.}
\label{fig09}
\end{figure}

We now treat the SCuP regime, in which the signals propagate in
opposite directions in the pump frame with the same slowness relative
to the pump pulse. Specifically, for this section we work with
parameter values: $\beta_s = 0$, $\beta_p = 2$, $\beta_r = 4$, $L =
1$, and Gaussian-shaped pump. For pump width $\tau_p=0.707$, the
low-CE Green function matches a time-shifted version of the plot in
Fig. \ref{fig02}. Increasing $\overline\gamma$ to higher-CE will
cause the selectivity $S$ to rise to a maximum, and then fall back to
lower values, just like in the SSVM regime, Fig. \ref{fig09} plots
selectivity vs. $\overline\gamma$ for various pump widths.

\begin{figure}[htb]
\centering
\includegraphics[width=5.2in]{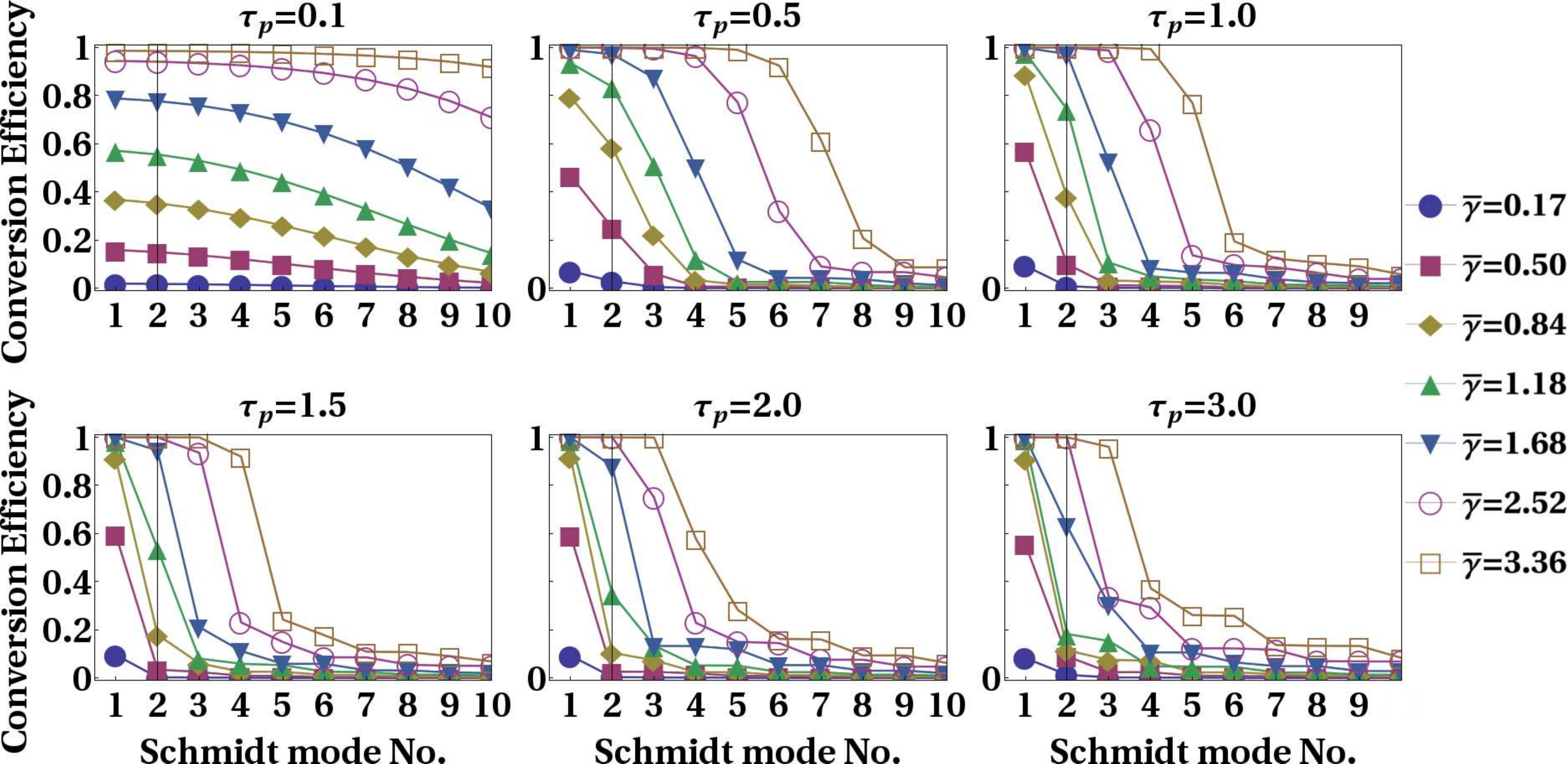}
\caption{Conversion efficiencies for the first ten Schmidt modes for
  various $\overline\gamma$ and Gaussian pump widths
  ($\tau_p$). $\beta_s = 0$, $\beta_p = 2$, $\beta_r = 4$, $L =
  1$. Numerical results.}
\label{fig10}
\end{figure}

We sought to improve this result by varying the low-CE GF aspect ratio
$(\tau_p/(\beta_{rs}L))$ by changing $\tau_p$. Starting from narrow
pumps and increasing the width, we could optimize the
selectivity-maxima to about $S\approx 0.7$ for a Gaussian pump width
of $\tau_p\approx 1.5$ and $\overline\gamma\approx 0.75$. Further
increasing $\tau_p$ stretched the GF shape in the $t=t'$ direction,
reducing its separability/selectivity-maximum, as shown in Fig.
\ref{fig09}. The selectivity maximum moves to larger $\overline\gamma$
for increasing $\tau_p$ because longer pump durations correspond to
smaller peak intensity.

Figure \ref{fig10} shows how the CEs for the first ten Schmidt modes
change with $\overline\gamma$ for various $\tau_p$. For large
$\overline\gamma$, higher-order CEs tend to decrease with increasing
$\tau_p$, suggesting mildly improved selectivity. They also appear to
oscillate about a decreasing central value in a damped fashion with
increasing $\tau_p$. For the values plotted, this is most pronounced
in the CE of the third Schmidt mode for $\overline\gamma=1.18$.

\begin{figure}[htb]
\centering
\includegraphics[width=3in]{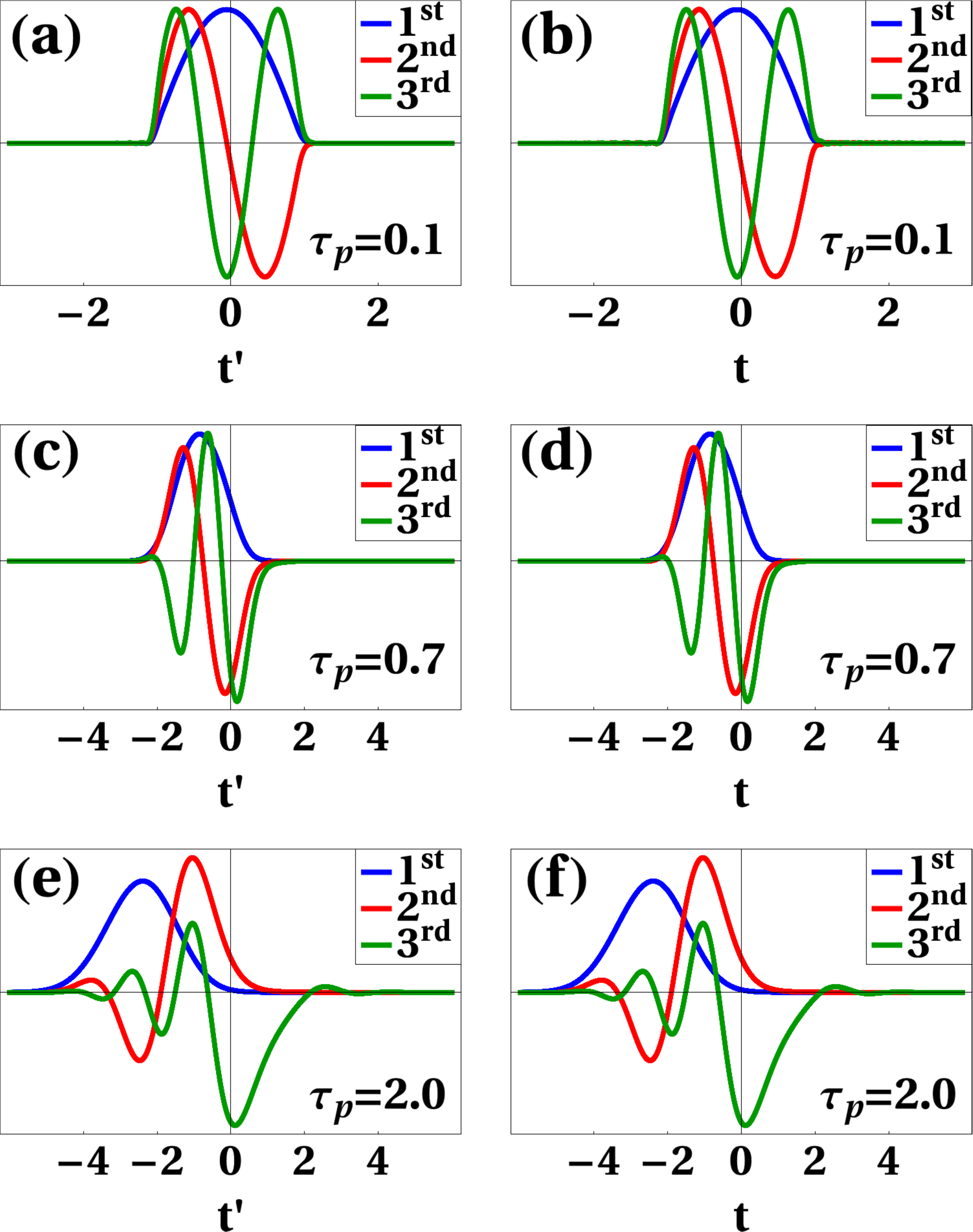}
\caption{The first three {\it s} input (a,c,e) and {\it r} output (b,d,f)
  Schmidt modes for $\overline\gamma=3.36$, and $\tau_p=$ $0.1$(a,b),
  $0.7$(c,d), and $2.0$(e,f). $\beta_s = 0$, $\beta_p = 2$, $\beta_r =
  4$, $L = 1$. Numerical results.}
\label{fig11}
\end{figure}

In this SCuP regime we find the shapes of the output (r) Schmidt modes
are essentially identical to those of the input (s) Schmidt
modes. Figure \ref{fig11} shows the dominant {\it s} input and {\it
  r} output Schmidt modes at $\overline\gamma=3.36$, for select
$\tau_p$. This shape-preserving behavior is related to the GF
consisting of the pump shape as a factor sloping parallel to the
$t=-t'$ direction, and is independent of $\overline\gamma$ for the
values tested. The individual Schmidt mode shapes, however, do change
with $\overline\gamma$. For $\tau_p\gg 0.1$, the Schmidt mode widths
scale linearly with the pump width.

\begin{figure}[htb]
\centering
\includegraphics[width=2in]{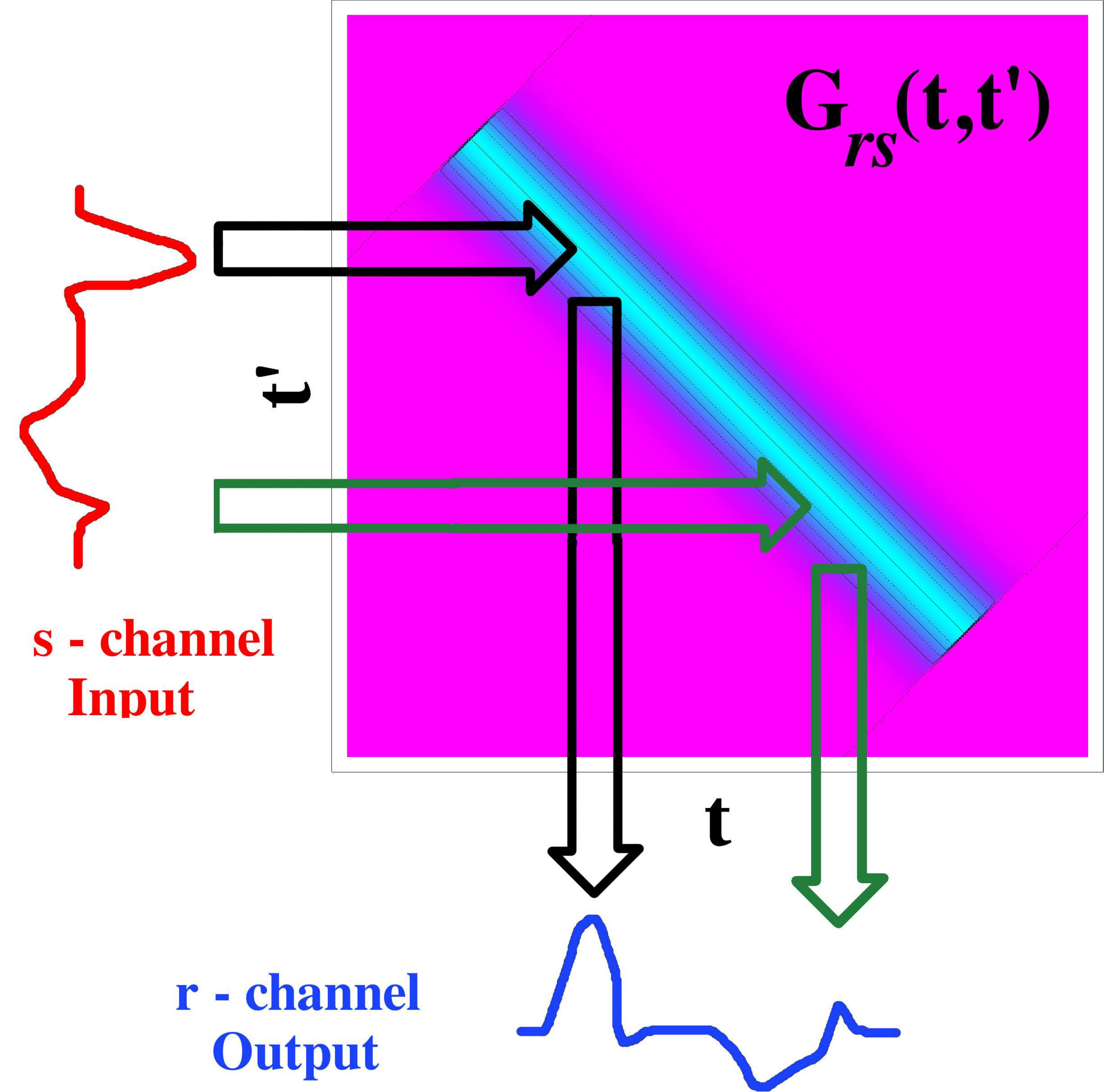}
\caption{Proposed mechanism for shape-preserving frequency conversion in
  the short-pump ``symmetrically counter-propagating signals'' regime.}
\label{fig12}
\end{figure}

The time-widths of the Schmidt modes have a lower bound of
$\beta_{rs}L/2$ due to the Heaviside step-function
boundaries. Decreasing $\tau_p$ to small values relative to
$\beta_{rs}L$ (e.g., $0.1$) causes the convergence of Schmidt mode
shapes to those plotted in Fig. \ref{fig11}(a,b). The dominant CE's
for the short-pump case nearly match each other in values, especially
for very low and very high $\overline\gamma$, making for a
non-selective add/drop device. These features cause the short-pump
SCuP regime to preserve the shapes of a large family of input pulses
during FC (even for CE's approaching unity). For example, for the
$\tau_p=0.1, \overline\gamma=3.36$ case in Fig. \ref{fig10}, all the
first seven Schmidt modes have near unity CE's. So any input pulse
that can be completely constructed by a linear superposition of the
first seven input Schmidt modes will FC into the other channel into
the exact same superposition of the first seven output Schmidt modes,
which also match the corresponding input Schmidt mode shapes (Fig.
\ref{fig11}). We hypothesize that this results from the $t'=-t$
direction of the GF being more pronounced for shorter $\tau_p$ (Fig.
\ref{fig12}), which maps local time-slices/segments of the input and
output pulses in a one-to-one fashion.

Note that the pump in the GF is constant along the $-45^\circ$
direction. So the interaction of each time segment of the signal with
the corresponding segment in the idler is driven by the same pump
profile. The resolution of such one-to-one segment mappings is
determined by the pump width. For broad pumps, any given time segment
of the signal would then influence a larger portion of the idler
pulse, and vice versa. The inability of the global shape of an input
pulse to influence its CE results in a poor add/drop device, but this
feature, which we call shape-preserving FC, has potential applications
in multi-color quantum interference \cite{enk10, huang92}.

\begin{figure}[htb]
\centering 
\includegraphics[width=3.2in]{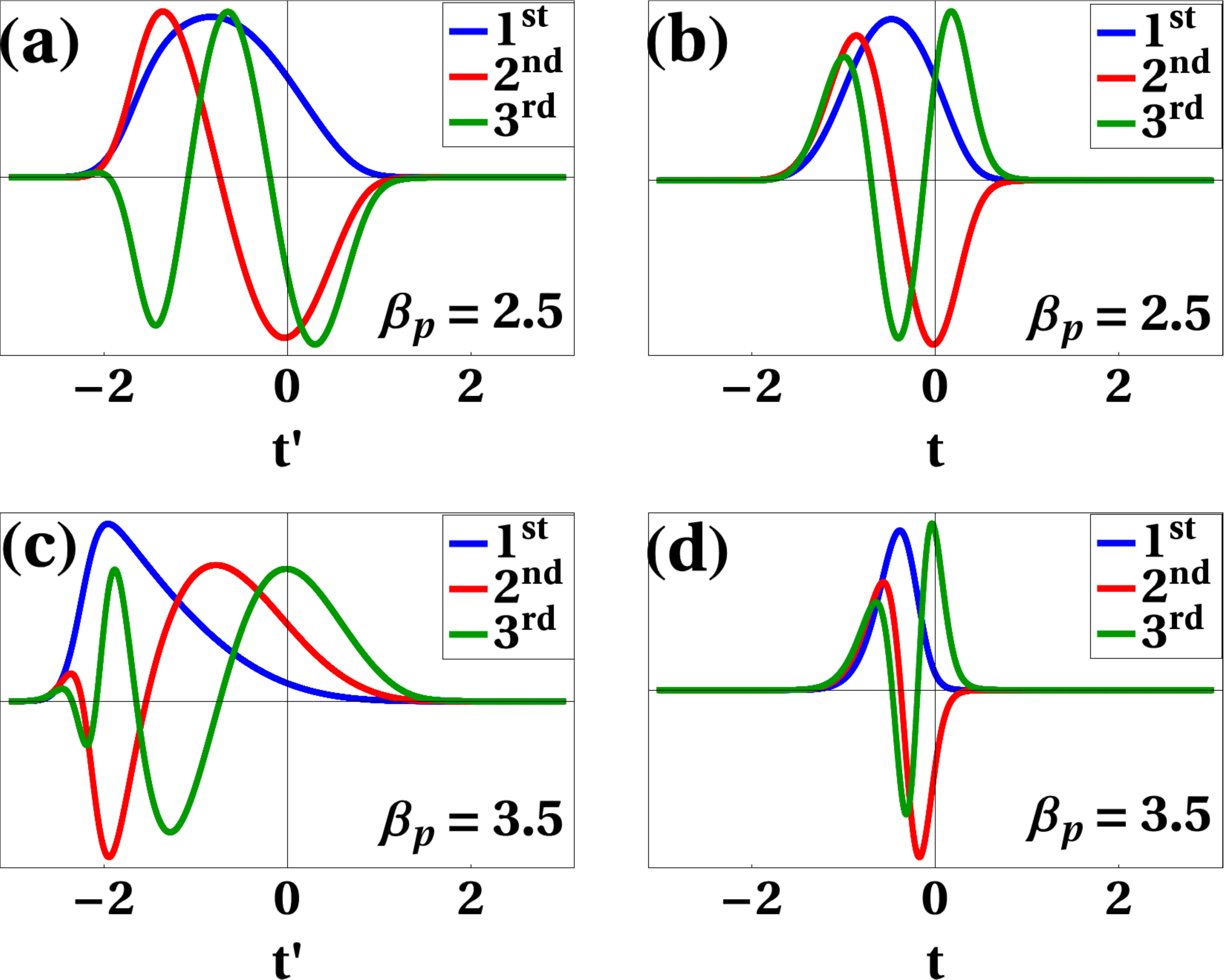}
\caption{The first three {\it s} input (a,c) and {\it r} output (b,d) Schmidt
  modes for $\overline\gamma=3.36$, $\tau_p=0.5$, and $\beta_p =
  2.5$(a,b), and $3.5$(c,d). $\beta_s=0$, $\beta_s=0$, $\beta_r=4$,
  $L=1$. Numerical results.}
\label{fig13}
\end{figure}
\subsection{($\beta_{rp}\beta_{sp}<0$) counter-propagating signals regime}

The previous section dealt with the parameter set $\beta_s=0$,
$\beta_r=4$, $\beta_p=2$. Holding the $\beta_s$ and $\beta_r$
slownesses at these values, we now vary the pump slowness ($\beta_p$)
within the range $[0,4]$ and chart the properties of the GF. At values
$0$ and $4$, the results matched those of the SSVM regime. The range
$\beta_p\in [0,2]$ showed a one-to-one symmetrically-mapped
correspondence with the range $[2,4]$. That is, for every $\Delta$ in
the range $[0,2]$, the GF had the same selectivities for
$\beta_p=(2-\Delta)$ as well as $(2+\Delta)$. Even the Schmidt modes
were identical but interchanged between the signal channels.

As the low-CE GF plots in Fig. \ref{fig01} show, the pump-shape
factor in $G_{rs}(t,t')$ has slope $\beta_{sp}/\beta_{rp}$, defined in
Eq. \eqref{13}. For fixed $L=1$, changing this slope, particularly for
small pump widths, will change the projected width of the GF on the
$t$ and $t'$ axes, which changes the widths 
\noindent of the Schmidt modes. This is also true for arbitrary CE's,
as is shown in Fig.  \ref{fig13}. Bringing $\beta_p$ closer to
$\beta_r$ will tend to align $G_{rs}(t,t')$ with the vertical
$t'$-direction. This increases the {\it s}-channel Schmidt mode
widths, and decreases the {\it r}-channel Schmidt mode widths.

\begin{figure}[htb]
\centering
\includegraphics[width=5.2in]{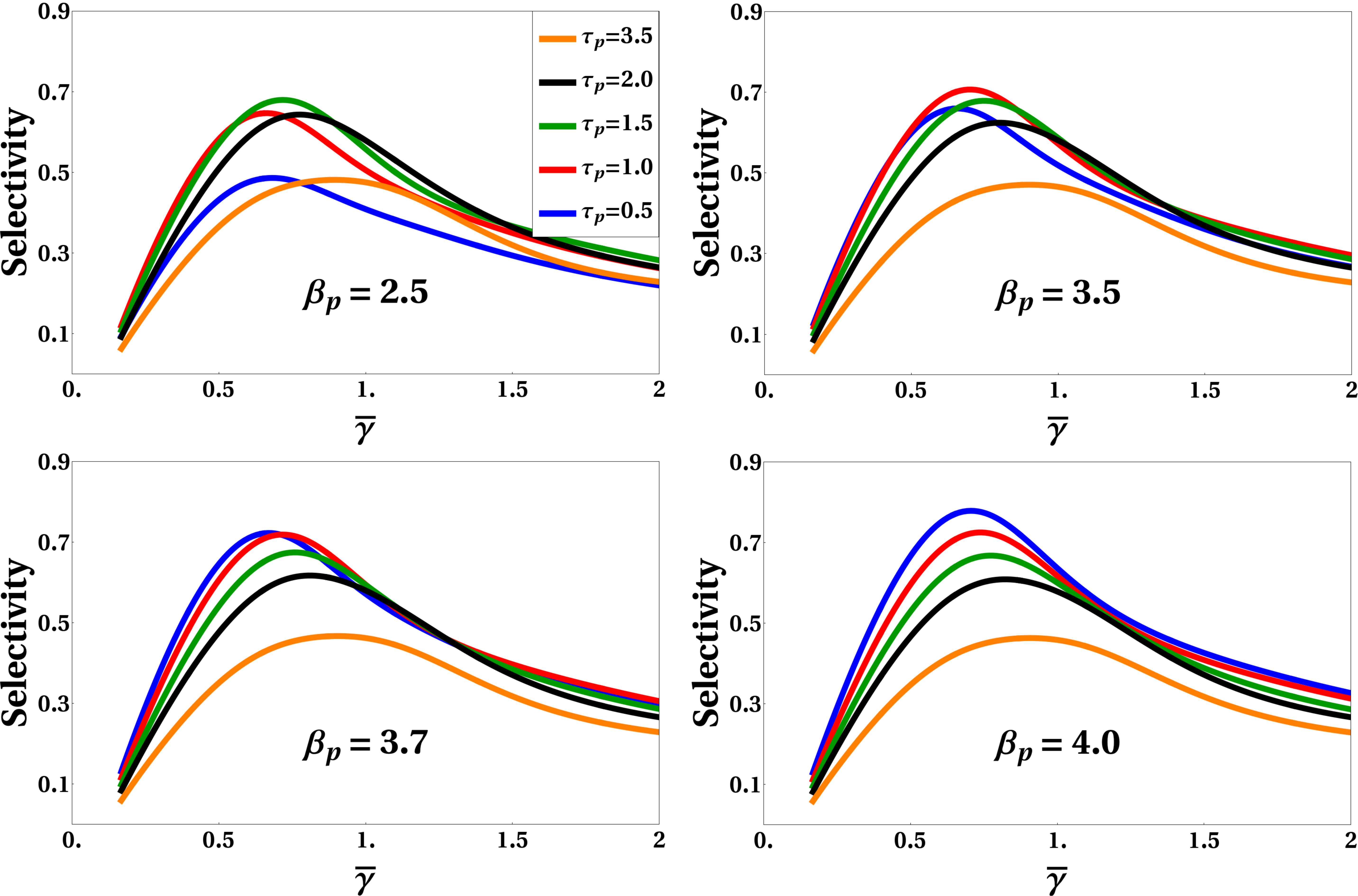}
\caption{Selectivity vs. $\overline\gamma$ for Gaussian pumps of
  various widths and various $\beta_p$. $\beta_s=0$, $\beta_r=4$,
  $L=1$. Numerical results.}
\label{fig14}
\end{figure}

Figure \ref{fig14} shows the plots of selectivity vs. $\overline\gamma$
for different $\beta_p$ values and pump widths. While in the SCuP
regime, the selectivity-maximum was highest for $\tau_p\approx 1.5$.
As $\beta_p$ drew closer to $\beta_r=4$, the optimum-selectivity-pump
width was seen to decrease. This is consistent with our finding for
the SSVM regime, which shows larger selectivity-maxima for shorter
pumps. The selectivity-maximum also increased as we approached the
SSVM regime.

The selectivities for short-pumps were hyper-sensitive to changes in
$\beta_p$ since the shape of the GF is affected the most (due to
pump-factor slope defined in Eq. \eqref{13}) for shorter pumps. This
implies that the closer we are to SSVM regimes (but not in it), the
shorter our pump needs to be for the FC to still be
shape-preserving. Selectitivies for wider-pumps did not show the same
sensitivity to changes in $\beta_p$.

%\clearpage

\begin{figure}[htb]
\centering \includegraphics[width=5.2in]{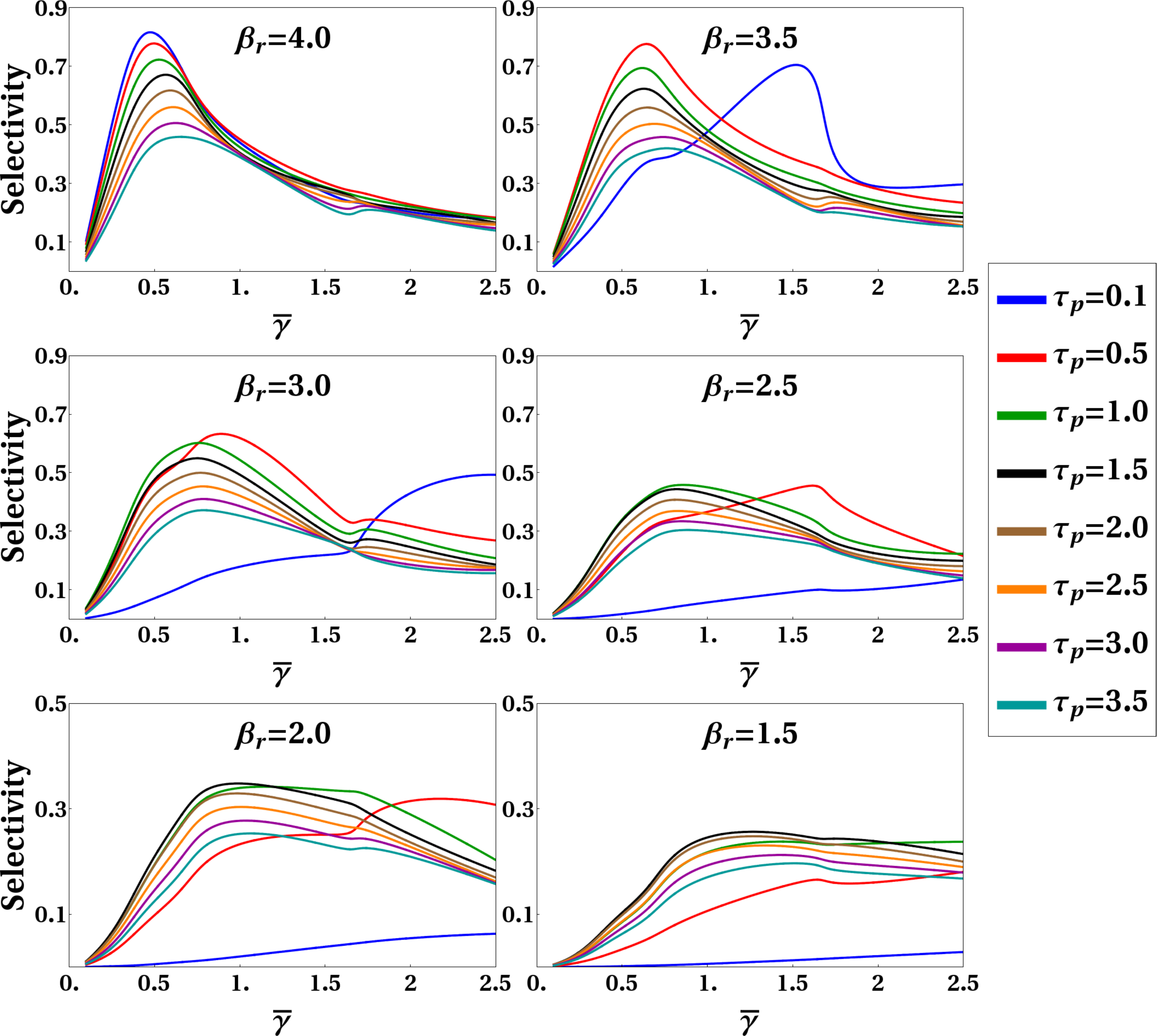}
\caption{Selectivity vs. $\overline\gamma$ for Gaussian pumps of
  various $\tau_p$ and $\beta_r$. $\beta_p=4$, $\beta_s=0$,
  $L=1$. Numerical results.}
\label{fig17}
\end{figure}

\subsection{($\beta_{rp}\beta_{sp}>0$) co-propagating signals regime}

In this section, we explore the regime in which the slope of the pump
factor in the low-CE GF, i.e. the quantity $\beta_{sp}/\beta_{rp}$ is
positive. We do this by fixing $\beta_p=4$, $\beta_s=0$, $L=1$, and
varying $\beta_r$ within the range $[0,4]$. Selectivity behavior for
negative values of $\beta_r$ mapped bijectively to the corresponding
positive $\beta_r$ that resulted in an inversion in pump-factor slope,
while the Schmidt modes swapped across the {\it r} and {\it s} channels.

\begin{figure}[htb]
\centering \includegraphics[width=3.4in]{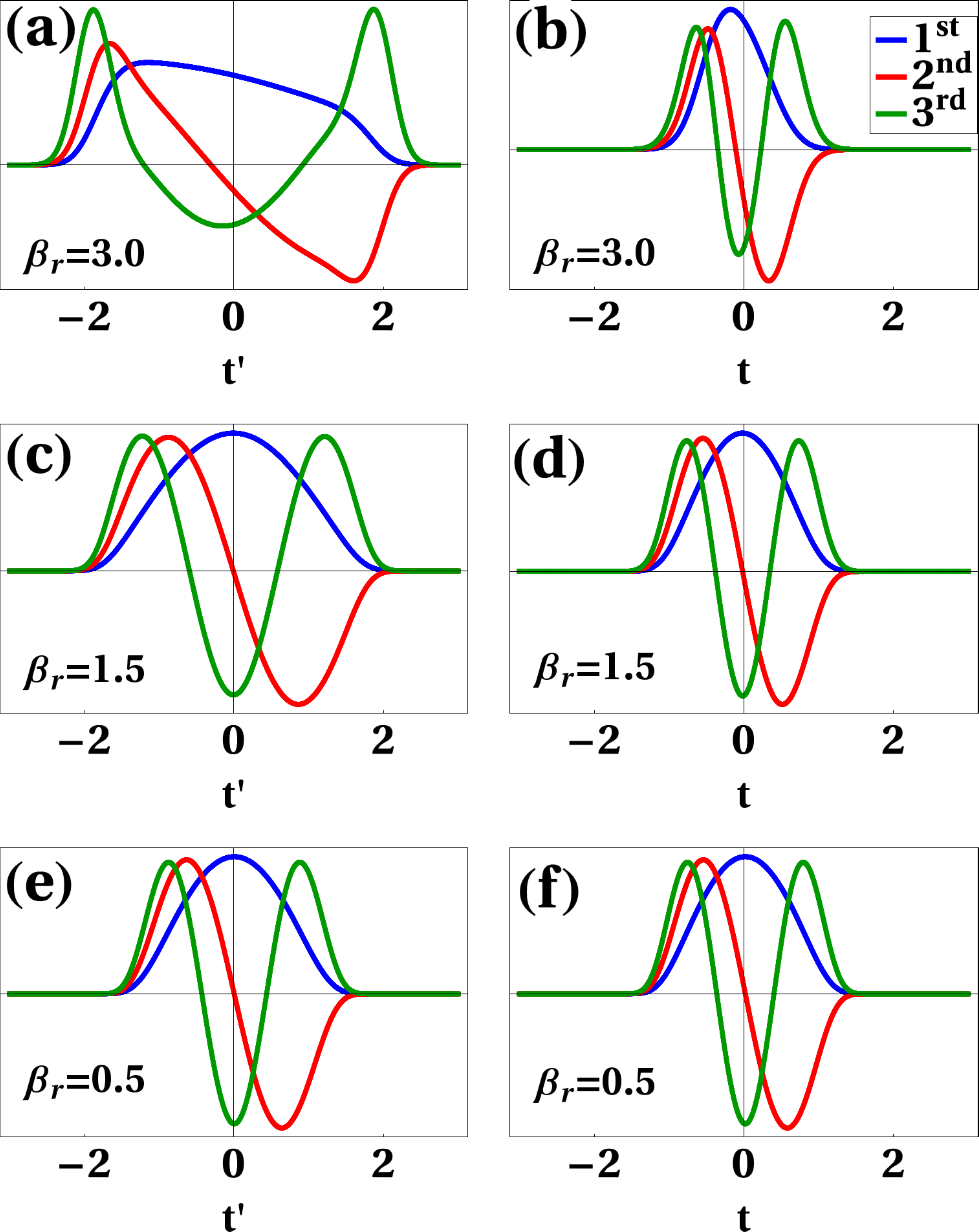}
\caption{The first three {\it s} input (a,c,e) and {\it r} output (b,d,f)
  Schmidt modes for $\overline\gamma=0.5$, $\tau_p=0.5$, and
  $\beta_r=3.0$(a,b), $1.5$(c,d), and $0.5$(e,f). $\beta_p=4$,
  $\beta_s=0$, $L=1$. Numerical results.}
\label{fig15}
\end{figure}

Figure \ref{fig17} consists of selectivity vs. $\overline\gamma$ plots
for various pump widths and $\beta_r$. The selecivity maximum for any
given $\tau_p$, apart from decreasing in magnitude with decreasing
$\beta_r$, also migrates to higher $\overline\gamma$ values. This
effect is more pronounced for shorter pumps. The optimum pump width
(with the highest selectivity maximum) also increases with decreasing
$\beta_r$.

As $\beta_r\to\beta_s$, the pump-factor slope approaches unity. This
allows for shape-preserving FC behavior when using short pumps,
through a mechanism analogous to that illustrated in Fig.
\ref{fig12}, except here the idler pulse convects through the pump in
the same direction as the signal pulse. CE's for the first ten Schmidt
modes for small $\beta_r$ tended to match each other, confirming
non-shape-descriminatory GF. This ``rotation'' of the GF pump-factor
causes the Schmidt mode widths to track the GF projection on the
$(t,t')$-axes. The difference is most noticeable for short pumps
(Fig. \ref{fig15}). The dominant Schmidt modes in both channels
converge to matching shapes, as expected for shape-preserving FC.

The pump-factor slope can also be made to approach unity by keeping
$\beta_r$ and $\beta_s$ fixed and increasing $\beta_p$ to very high
magnitudes. This approach would maintain the spacing between the
Heaviside step-functions and prevent the selectivity maximum from
migrating to higher $\overline\gamma$ values. Numerical constraints
restrain us from covering the entire range of the pump-factor slope
using this method.

\subsection{($\beta_r=\beta_s$) Exactly co-propagating signals regime}

The ECoP regime is special in that we cannot plot the low-CE GF as we
did for all the other regimes. As $\beta_{rs}\rightarrow 0$, the
separation between the Heaviside step-functions also converges to
zero. We can however, explicitly write down the complete analytical
solution for real pump-functions. If $\beta_s=\beta_r=0$ and
$A_p(x)\in\mathbb{R}$, then:

\begin{align*}
\partial_zA_r(z,t)& = i\gamma A_p(t-\beta_pz) A_s(z,t),\tag{15a}\label{15a}\\
\partial_zA_s(z,t)& = i\gamma A_p(t-\beta_pz) A_r(z,t),\tag{15b}\label{15b}\\
A_r(L,t)=&A_r(0,t)\cos\left[P(L)\right]+iA_s(0,t)\sin\left[P(L)\right],\tag{16a}\label{16a}\\
A_s(L,t)=&A_s(0,t)\cos\left[P(L)\right]+iA_r(0,t)\sin\left[P(L)\right],\tag{16b}\label{16b}
\end{align*}

\noindent where $P(z):=(\gamma/\beta_p)\int^t_{t-\beta_pz}A_p(x)dx$,
and $\lim_{\beta_p\to 0}P(z) = \gamma A_p(t) z$.  The GF are
$\delta$-functions in $t'$, and do not lend themselves to numerical
Schmidt decomposition. This regime is beyond the scope of our
simulation methodology (detailed in Appendix I).

The absence of walk-off between the two signal channels implies that
the evolution of $A_r(z,t)$ for a given local time index `$t$' is
insensitive to the global shapes of the input wavepackets
($A_r(0,t')$, $A_s(0,t')$). This results in a poor
add/drop-device. Different time slices of arbitrary input pulses will
undergo the same transformation as they sweep across the pump,
allowing for distortionless conversion.

The SSVM regime ($\beta_s=\beta_p$), under the $\beta_{rs}\to 0$ limit
also converges to Eqs. \eqref{16a} and \eqref{16b} (as verified in
Appendix II). The exact solution in this regime for arbitrary
complex-valued (chirped) pump functions will not be dealt with in this
publication.

\section{Analytical solution for single sideband velocity matched regime}
The SSVM regime, where $\beta_s=\beta_p$, and all other parameters are
arbitrary, was shown above to be the optimal regime for the drop/add
process. Fortunately, in this same regime the problem can be solved
analytically, following \cite{col12}(detailed in Appendix III). The
exact GF is found to be
\begin{align*}
 G_{rr}&(t,t') = H(\tau - \tau')\delta(\zeta-\zeta') - \overline\gamma\sqrt{\eta/\xi}J_1\{2\overline\gamma\sqrt{\eta\xi}\} H_H(\tau,\tau',\zeta,\zeta'),\tag{17a}\label{17a}\\
G_{sr}&(t,t') = i\overline\gamma A_p^*(\tau)J_0\{2\overline\gamma\sqrt{\eta\xi}\} H_H(\tau,\tau',\zeta,\zeta'),\tag{17b}\label{17b}\\
G_{rs}&(t,t') =i\overline\gamma A_p(\tau')J_0\{2\overline\gamma\sqrt{\eta\xi}\}H_H(\tau,\tau',\zeta,\zeta'),\tag{17c}\label{17c}\\
G_{ss}&(t,t') =\delta(\tau - \tau')H(\zeta-\zeta') - \overline\gamma A_p^*(\tau)A_p(\tau')\sqrt{\zeta/\eta} J_1\{2\overline\gamma\sqrt{\eta\xi}\} H_H(\tau,\tau',\zeta,\zeta').\tag{17d}\label{17d}
\end{align*}
\noindent Here $\tau = t-\beta_sL$, $\tau' = t'$, $\zeta =
\beta_rL-t$, $\zeta' = -t'$, $\xi=\zeta-\zeta'$, $\overline\gamma =
\gamma/\beta_{rs}$, and $\eta =
\textstyle\int^\tau_{\tau'}|A_p(x)|^2dx$. \\ $J_n(...)$ is the Bessel
function of order $n$, and $H_H(\tau,\tau',\zeta,\zeta')=H(\tau -
\tau')H(\zeta-\zeta')$, $H(x)$ being the Heaviside step-function.

\begin{figure}[htb]
\centering
\includegraphics[width=3.2in]{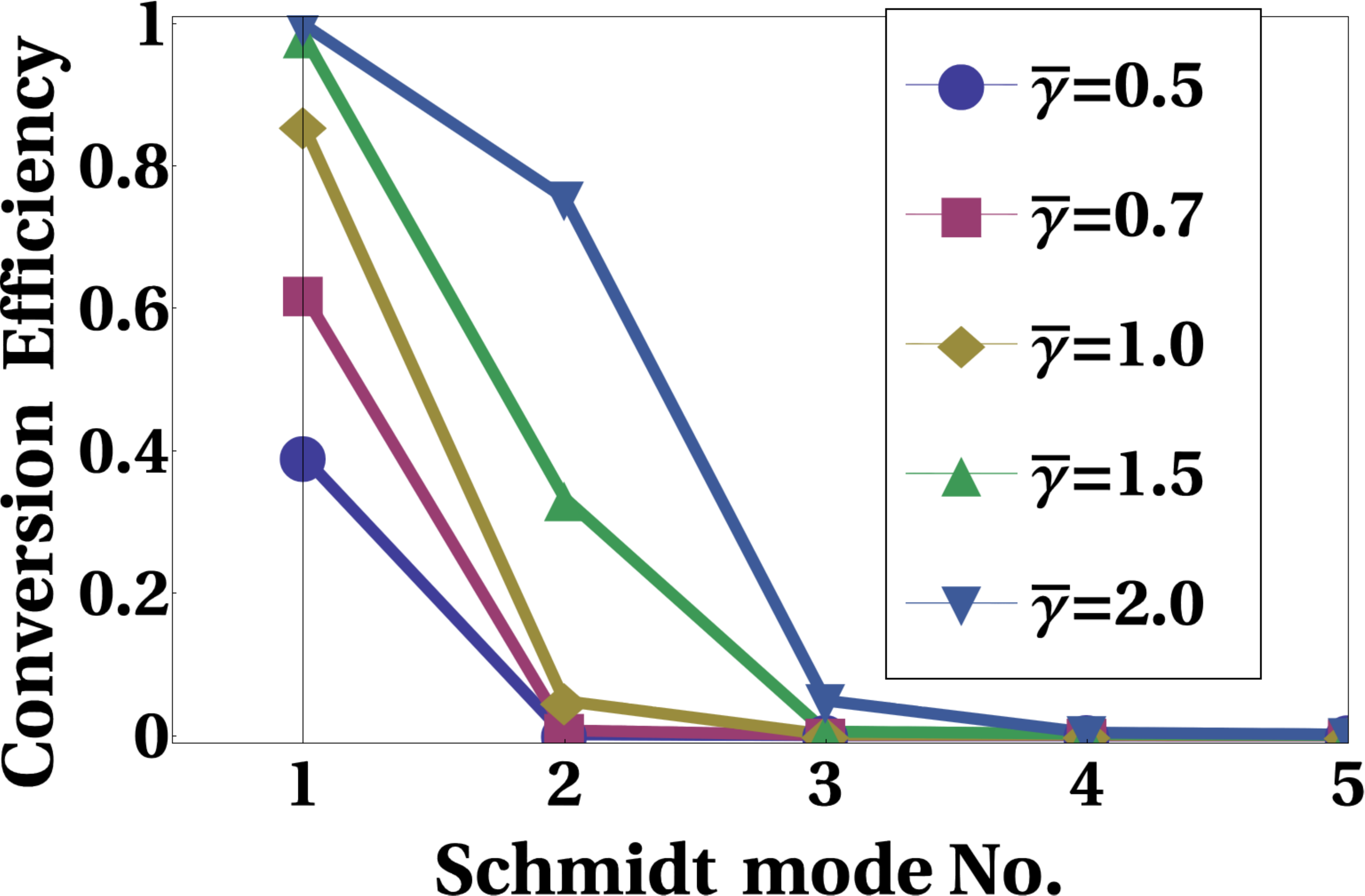}
\caption{The first five dominant conversion efficiencies for the
  parameters in Fig. \ref{fig03}(b) and \ref{fig06}, for various
  $\overline\gamma$, SSVM regime. Derived via SVD of the exact Green
  function $G_{rs}(t,t')$ (Eq. \eqref{17c}).}
\label{figx2}
\end{figure}
For an analysis of selectivity/separability, we need only consider the
structure of $G_{rs}(t,t')$, which has two non-separable factors in
$(t,t')$: the Bessel function $J_0\{2\overline\gamma\sqrt{\eta\xi}\}$, and the
step-functions $H_H(\tau,\tau',\zeta,\zeta')$. Decreasing the pump
width relative to the effective interaction time ($\beta_{rs}L$) can
diminish the ill-effects of the step-functions on GF-separability, but
the effect of the Bessel function worsens at higher
$\overline\gamma$. A numerical singular value decomposition of this
analytical GF in Eq. \eqref{17c} for high $\overline\gamma$ plotted in
Fig. \ref{figx2} confirms our numerical results from section 4.1.

\begin{figure}[htb]
 \centering
\includegraphics[width=2.7in]{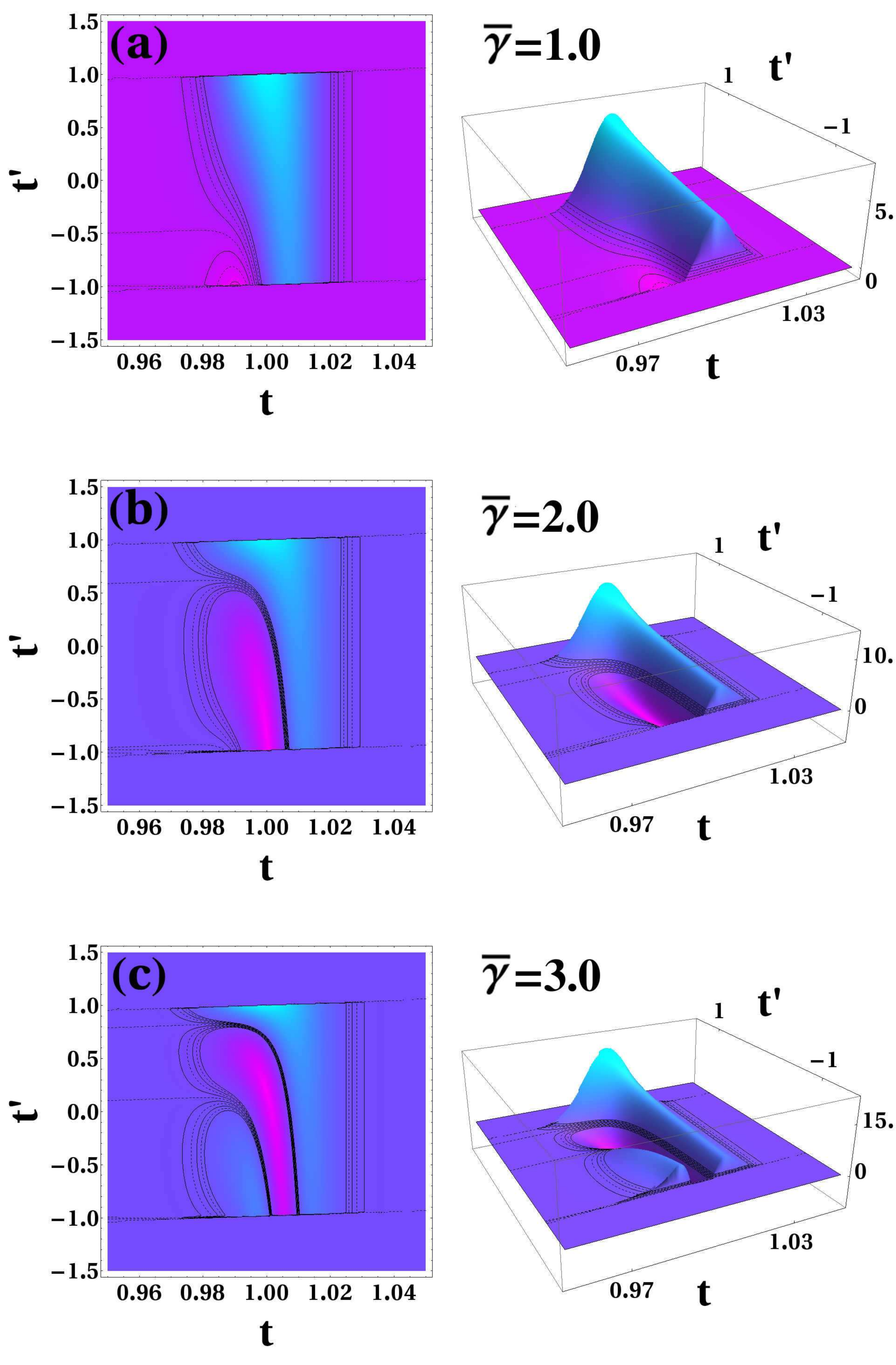}
\caption{Green function for parameters in Fig. \ref{fig03}(a), $\tau_p
  = 0.01$, top and perspective-views, for {\bf (a)} $\overline\gamma =
  1.0$, {\bf (b)} $\overline\gamma = 2.0$, {\bf (c)} $\overline\gamma
  = 3.0$.}
\label{figx3}
\end{figure}

\begin{figure}[htb]
\centering
\includegraphics[width=3.6in]{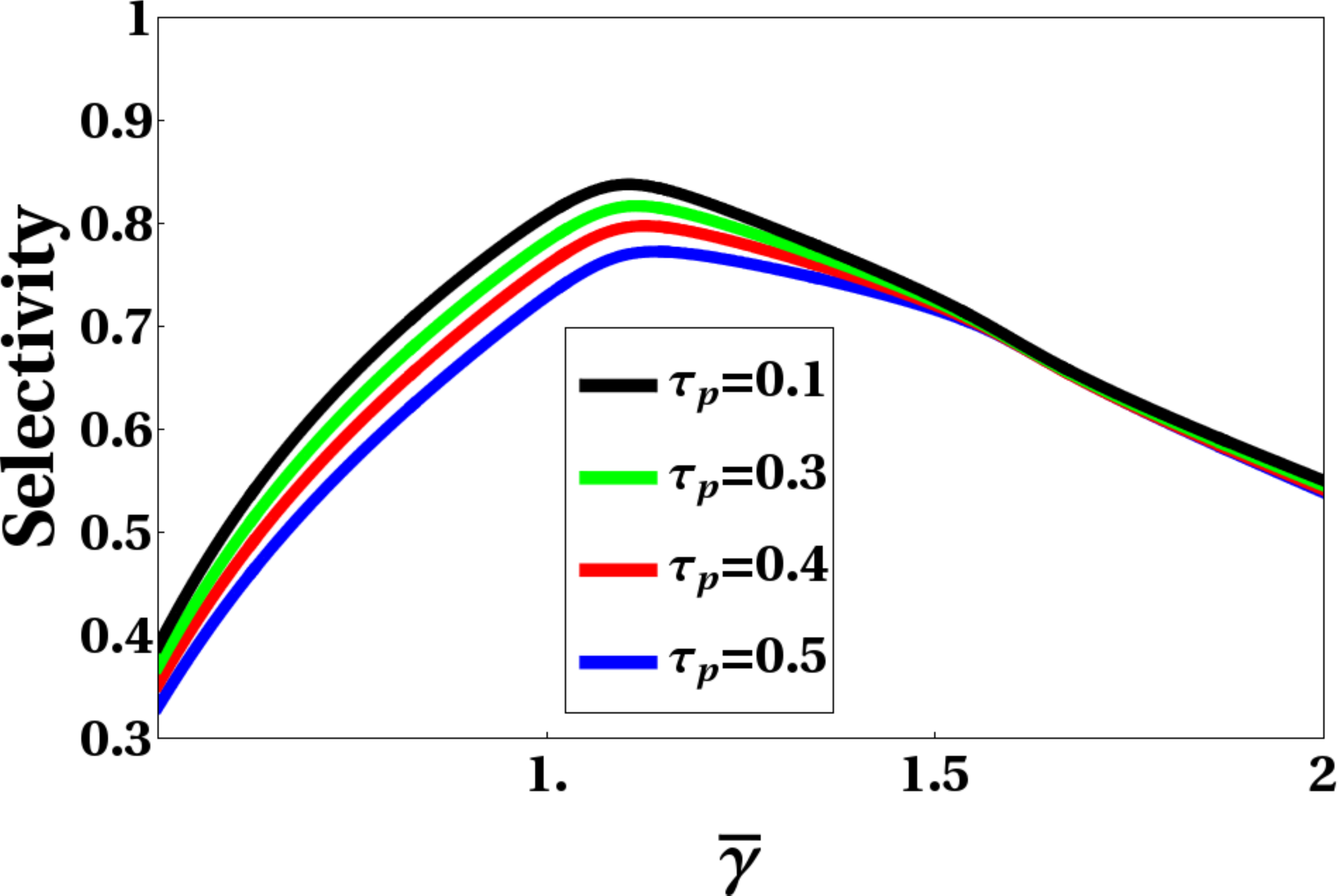}
\caption{Selectivity vs. $\overline\gamma$ for parameters from Fig.
  \ref{fig03}(b) and \ref{fig06}, for various pump widths ($\tau_p$),
  using $G_{rs}(t,t')$ in Eq. \eqref{17c}.}
\label{figx1}
\end{figure}

Increasing $\overline\gamma$ improves the CE of the first Schmidt mode
by scaling the peak of the GF, but via the Bessel function, decreases
the separability (Fig. \ref{figx3}).  Hence, selectivity, being a
product of the two, attains a maximum value at around $\overline\gamma
\approx 1.15$. While decreasing pump width ($\tau_p$) improves
selectivity, the maximum asymptotically approaches a limiting value of
approximately $0.85$ (Fig. \ref{figx1}). This Bessel function
induced distortion in GF shape is reflected in the shape of the
Schmidt modes (section 4.1).


One might suspect that including a frequency chirp in the
pump field could improve the selectivity. We prove here that for the
SSVM regime this is not the case. Note that the pump-squared integral
$\eta(t,t')$, and consequently the Bessel function, is independent of
any pump-chirp. To demostrate this, we rewrite Eqs. \eqref{1a} and
\eqref{1b} in the pump's moving frame:
\begin{align*}
(\partial_z+\beta_{rp}\partial_t)A_r(z,t) = i\gamma A_p(t) A_s(z,t),\tag{18a}\label{18a}\\
(\partial_z+\beta_{sp}\partial_t)A_s(z,t) = i\gamma^* A_p^*(t) A_r(z,t).\tag{18b}\label{18b}
\end{align*}

Replacing the pump envelope-function by its real-amplitude and phase
($A_p(t):=P(t)\exp[i\theta(t)]$) and setting $\beta_{sp}=0$ for the
SSVM constraint, we get:
\begin{align*}
(\partial_z+\beta_{rp}\partial_t)A_r(z,t) &= i\gamma P(t)\exp[i\theta(t)] A_s(z,t),\tag{19a}\label{19a}\\
\partial_zA_s(z,t) &= i\gamma^* P(t)\exp[-i\theta(t)] A_r(z,t).\tag{19b}\label{19b}
\end{align*}
By redefining the {\it s}-channel envelope function as
$\overline{A}_s(z,t) = A_s(z,t)\exp[i\theta(t)]$, we can recover
Eqs. \eqref{1a} and \eqref{1b} with a real-pump envelope in the SSVM
regime. Any time dependent complex phase in the pump gets absorbed
into the Schmidt modes, without affecting the CE's or GF
selectivity. Nevertheless, for any given $\overline\gamma$, the shape
of the pump gives us some control over the shapes of the Schmidt
modes, and this may be used to tune the add/drop device to accept
easy-to-produce pulse shapes as input Schmidt modes.

\begin{figure}[tbh]
\centering
\includegraphics[width=3.6in]{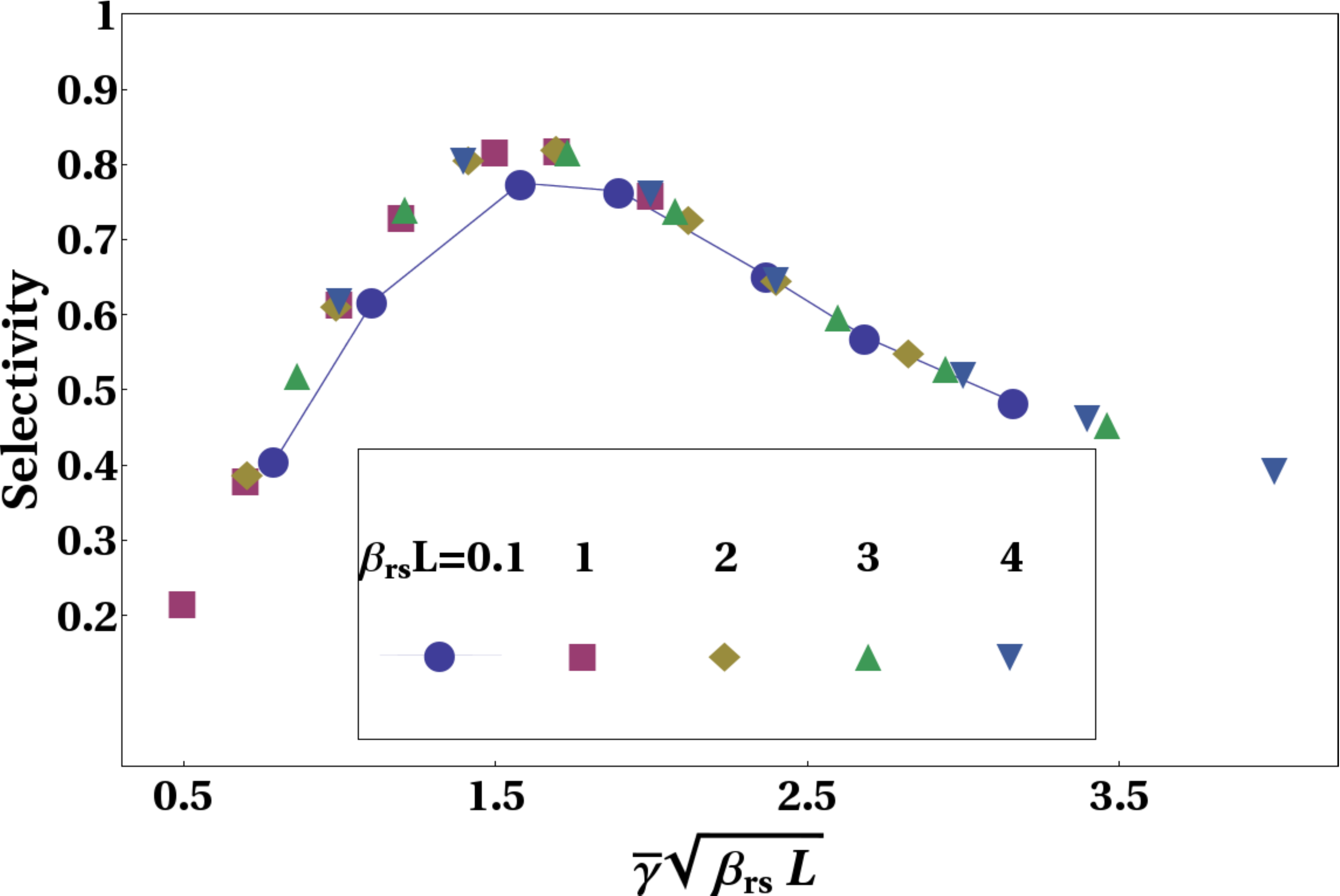}
\caption{Selectivity vs. $\overline\gamma\sqrt{\beta_{rs}L}$ for
  parameters from Fig. \ref{fig03}(b), with various
  $\beta_{rs}L$. The joined plot for $\beta_{rs}L=0.1=\tau_p$ has a
  lower maximum than all other plots.}
\label{figx4}
\end{figure}

The parameter ($\beta_{rs}L$) is responsible for the Schmidt mode
width for the channel with velocity mismatched with that of the
pump. This parameter has units of time, and is a measure of the
duration of ``interaction''. Increasing $\beta_{rs}L$ in the SSVM
regime will make higher CE's attainable at higher pump powers but
lower $\overline\gamma$. The selectivity maximum also follows a
similar trend until $\beta_{rs}L$ becomes comparable to pump width
$\tau_p$ (at which point the slope of the Heaviside step-functions
reduces overall GF separability), as shown in Fig. \ref{figx4}.

The non-separability arising from the Bessel function in the GF can be
traced to the oscillations shown in Fig. \ref{figx3}. These are
similar to those in Burnham-Chiao ringing \cite{Burnham1969} seen in
fluorescence induced by short-pulse excitations by the propagation of
short, weak pulses through a resonant atomic medium. To model an
analogy, the phase- and energy-matched wave-mixing process may be
represented by a 2-level pseudo-atomic-medium with a ground-state
energy at $\omega_p$ and an excited state at $\omega_s$. Any
finite-width input pulse in {\it r} channel with energy resonant with the
atomic-medium ($\omega_r = \omega_s-\omega_p$) will have a non-zero
bandwidth in the frequency domain. As it interacts with the medium,
its spectral-components detuned above resonance will acquire a
different phase shift than the spectral components below
resonance. These two spectral components will beat to produce the
ringing effect, resulting in the oscillations seen in Fig.
\ref{figx3}. The Bessel function factor is a fundamental barrier that
restricts selectivity in the SSVM regime (which is thought to be the
optimal one). Some groups \cite{silb11, bran11} have sought to make the
waveguide properties non-uniform ($\gamma\rightarrow\gamma(z)$) in an
attempt to overcome this limitation, with limited success. A full
analysis of that regime is beyond the scope of this publication.

\section{Summary and concluding remarks}

In this paper, we modelled sum/difference frequency generation
processes involving three-wave mixing with a strong pump in a uniform
but finite $\chi^{(2)}$-nonlinear medium. We used the notion of Green
function separability for such processes and used it to study the
feasibility of using these processes for implementing an orthogonal
time-frequency-division multiplexer (OTFDM). Such a device would
selectively discriminate between orthogonal weak classical pulses or
single-photon wavepackets that overlap both in temporal and frequency
domains. We employed singular-value-decomposition (or Schmidt
decomposition) of said Green functions to define selectivity: a figure
of merit that quantifies the process's viability for application as an
OTFDM add/drop device. The decomposition also produced
parameter-dependent input and output Schmidt modes, which functioned
as a natural orthogonal basis set of channel waveforms for the device
to multiplex.

Under perfect phase-matching conditions, and ignoring higher-order
dispersion (valid for long-enough pulses) as well as self- and
cross-phase modulation effects, we identified the group slownesses of
the participating optical channels, the medium length, the pump
pulse-shape and pump power as all the parameters that determine the
Green function. We were able to perturbatively approximate the Green
functions for low pump powers and contrast the separabilities
associated with various configurations and parameter regimes, subject
to the above mentioned constraints. We then undertook an exhaustive
numerical computation of selectivities [defined in Eq. \eqref{5}] in
all regimes for real chirp-free Gaussian pump-pulses of arbitrary
power and width within the above mentioned constraints. 

We found that the best selectivity ($\sim 0.83$) is obtained when the
group slowness of one of the signal channels is matched with that of
the pump, a regime first proposed in \cite{bib02}, and discussed in
sections 4.1 and 5. This SSVM regime resulted in Schmidt mode widths
that equaled the pump width for the group-slowness-matched signal
channel, and equaled the effective inter-pulse-interaction time
($\beta_{rs}L$) for the other channel. We then presented the complete
analytical solution for this regime, which sheds light on certain
parameter-scale invariances, whilst imposing a strict upper bound on
the selectivity and proving its independence of pump-chirping.

We also found that symmetrically mismatching the group
slownesses of the signal channels, making one of them slower than the
pump and the other one faster, also yielded reasonable selectivity (up
to $0.7$) for moderate pump widths. This regime also had Schmidt modes
of equal time widths in both channels, and for short pumps resulted in
pulse shape-preserving frequency conversion with high efficiency.

We conjecture that further boosts to selectivity can be acheived only
by the introduction of non-uniformity in the nonlinear medium,
introduction of higher-order dispersion, or reliance on higher order
processes such as four-wave-mixing.

{\bf Note added in proof:} A very recent related publication proposed
that approximately single-mode QFC can be obtained independently of
the phase-matching regime by choosing a pump pulse duration comparable
to and/or shorter than the inverse of the phase-matching
bandwidth\cite{huang13}. In the present study we did not find evidence
that high selectivity, as defined by our figure of merit
Eq. \eqref{5}, can be obtained generally in this manner. We find that
the SSVM regime, proposed by Eckstein \cite{bib02}, results in
significantly higher selectivity than all other regimes.

The work of DR, MR and CM was supported by the National Science
Foundation through ECCS and GOALI. We thank Dr.~Craig Rasmussen for
his help with parallelization of the simulation code. All numerical
computations presented in this paper were performed on the ACISS
cluster at the University of Oregon.

\section*{Appendix I}
{\bf Procedure for numerical derivation of Green function}

In order to derive the numerical GF for TWM, we first implemented a
coupled mode equation solver that accepts arbitrary input functions
($A_r(0,t')$, $A_s(0,t')$) as arguments, and computes the resultant
output functions ($A_r(L,t)$, $A_s(L,t)$) for Eqs.  \eqref{1a} and
\eqref{1b}. This is achieved using a Runge-Kutta based method. The
solver iterates over differential steps in pulse-propagation ($\Delta
z$) from $z = 0$ to $L$ (medium length). Every iteration consists of
an upwinded $z$-propagation scheme for all three pulses (signal, pump
and idler) by a step ($\Delta z$), followed by a Runge-Kutta
implementation of the coupled nonlinear interaction, all in spacetime
domain.  Next, we compute the GF by computing the outputs for an
orthogonal set of input `test signals'. To ellaborate, consider the GF
submatrix and its singular-value-decomposition (SVD):
\begin{equation*}
G_{rs}(t,t') = \sum\limits_j\rho_j\Psi_{j}(t)\phi^*_{n}(t').\tag{20}\label{20}
\end{equation*}
\noindent The objective is to calculate all the individual components
($\rho_j$, $\Psi_{j}(t)$ , $\phi_{n}(t')$) on the
right-hand-side. We first pick a spanning-set of basis functions
$\{B_{r,k}\}$ and $\{B_{s,l}\}$ and re-express the Schmidt modes:
\begin{align*}
&\Psi_{j}(t)=\sum\limits_kU_{jk}B_{r,k}(t);\quad\phi^*_{j}(t')=\sum\limits_lV_{jl}B^*_{s,l}(t'),\tag{21a}\label{21a}\\
G_{rs}(t,t')=&\sum\limits_{k,l}\left[\sum\limits_jU_{jk}\rho_jV_{jl}\right]B_{r,k}(t)B^*_{s,l}(t')=\sum\limits_{k,l}\left[\overline{G}_{rs}\right]_{kl}B_{r,k}(t)B^*_{s,l}(t').\tag{21b}\label{21b}
\end{align*}
\noindent Using ($A_r(0,t)=0$, $A_s(0,t')=B_{s,l}(t')$) as inputs for
the solver, and decomposing the resulting {\it r}-channel outputs
$A_r(L,t)$ in the $\{B_{r,k}\}$ basis will yield the entire
$l^{th}$-column of the complex matrix $\overline{G}_{rs}$. Once this
matrix is determined, its SVD will directly reveal $\{U_{jk}\}$,
$\{V_{jl}\}$, and $\{\rho_j\}$, and through them, the Schmidt
modes. For the results presented in this publication, we chose
Hermite-Gaussian functions for the spanning-set of basis functions for
both input and output Schmidt modes, since the low-CE Schmidt modes
for Gaussian pump shapes are nearly Hermite-Gaussian.

\section*{Appendix II}
{\bf SSVM solution converging to ECoP in the
  $\beta_{rs}\rightarrow 0$ limit }

Taking the $\beta_{rs}\rightarrow 0$ limit whilst still enforcing SSVM
conditions will cause $\beta_p \rightarrow 0$ as well. We now verify
that the exact SSVM analytical solution consistently reduces to the
expected sinusoidal form expressed in section 4.5. Consider
$G_{rs}(t,t')$ from Eq. \eqref{17c} for the input condition
$A_r(0,t')=0$:
\begin{equation*}
A_r(L,t)=i\overline\gamma\int\limits^{t-\beta_sL}_{t-\beta_rL}dt'A_p(t')J_0\{2\overline\gamma\sqrt{\eta\xi}\}A_s(0,t')\tag{22a}\label{22a}
\end{equation*}
\begin{equation*}
=i\frac{\gamma
  L}{\beta_{rs}L}\int\limits_0^{\beta_{rs}L}dt''A_p(t''+t-\beta_rL)A_s(0,t''+t-\beta_rL)J_0\left[\frac{2\gamma
  L}{\beta_{rs}L}\left(t''\int\limits^{t-\beta_rL+\beta_{rs}L}_{t''+t-\beta_rL}|A_p(x)|^2dx\right)^{1/2}\right].\tag{22b}\label{22b}
\end{equation*}
\noindent Since $t''$ is being integrated from $0$ to $\beta_{rs}L$,
$\beta_{rs}\to 0 \Rightarrow t''\to 0$. Then the integral inside the
bessel argument reduces to
$(\beta_{rs}L-t'')|A_p(t-\beta_rL)|^2$.\\ 
We then use:
\begin{align*}
\frac{g}{y}\int\limits^y_0dt''J_0\left[|g|\frac{\sqrt{t''(y-t'')}}{y}\right]=2\sin\left(\frac{g}{2}\right)\tag{23a}\label{23a}\\
\Rightarrow A_r(L,t)=iA_s(t-\beta_rL)\sin\left[\gamma LA_p(t-\beta_rL)\right],\tag{23b}\label{23b}
\end{align*}
\noindent which is identical to Eq. \eqref{16a} for $\beta_r=0$ and
$A_r(0,t)=0$.

\section*{Appendix III}
{\bf Analytical derivation of the TWM Green function in the
  single sideband group-velocity matched regime}

Consider the equations of motion (Eqs. \eqref{1a} and \eqref{1b}) for
frequency conversion (FC) by three-wave mixing (TWM):
\begin{align*}
 (\partial_z + \beta_r\partial_t)\bar{A}_r(z,t) = i\gamma A_p(t - \beta_pz)\bar{A}_s(z,t),\tag{24a}\label{24a}\\
(\partial_z + \beta_s\partial_t)\bar{A}_s(z,t) = i\gamma A_p^*(t - \beta_pz)\bar{A}_r(z,t).\tag{24b}\label{24b}
\end{align*}
\noindent For the group-velocity matched case in which $\beta_s =
\beta_p$, it is convenient to define the retarded time variable $\tau
= t - \beta_sz$ and the normalized distance variable $\zeta = \beta_rz
- t$. By using these variables, one can rewrite the equations of
motion in the simplified forms
\begin{align*}
\partial_\tau A_r(\tau,\zeta) &= i\overline\gamma A_p(\tau)A_s(\tau,\zeta),\tag{25a}\label{25a}\\
\partial_\zeta A_s(\tau,\zeta) &= i\overline\gamma A_p^*(\tau)A_r(\tau,\zeta),\tag{25b}\label{25b} 
\end{align*}
\noindent where the modified coupling coefficient $\overline\gamma =
\gamma/\beta_{rs}$ and the differential slowness (walk-off parameter)
$\beta_{rs} = \beta_r - \beta_s > 0$. The equations are to be solved
for $-\infty < \tau < \infty$ and $0 \le \zeta < \infty$. An easy way
to do this is by Laplace transformation in space ($\zeta \rightarrow
s$) \cite{ray81,gia95}. If an impulse is applied to the signal at the
input boundary ($z = 0$), the transformed equations are
\begin{align*}
\partial_\tau A'_r(\tau,s) &= i\overline\gamma A_p(\tau)A'_s(\tau,s),\tag{26a}\label{26a}\\
s A'_s(\tau,s) &= i\overline\gamma A_p^*(\tau)A'_r(\tau,s) + \delta(\tau - \tau'),\tag{26b}\label{26b}
\end{align*}
\noindent where $\tau'$ is the  source time. If $H(\tau-\tau')$ is the
Heaviside   step   function   and    the   effective   time   variable
$\eta(\tau,\tau')  = \textstyle\int_{\tau'}^\tau  |A_p(x)|^2dx$, the  solutions
are:
\begin{align*}
A'_r(\tau,s) =& i\overline\gamma \frac{A_p(\tau')}{s} \exp[-\overline\gamma^2\eta(\tau,\tau')/s] H(\tau - \tau'),\tag{27a}\label{27a}\\
A'_s(\tau,s) =& \frac{\delta(\tau - \tau')}{s} - \left[\frac{\overline\gamma^2A_p^*(\tau)A_p(\tau')}{s^2}\right]\exp[-\overline\gamma^2\eta(\tau,\tau')/s] H(\tau - \tau').\tag{27b}\label{27b}
\end{align*}
One can rewrite solutions in the space domain by using tables of
inverse transforms \cite{chu72}. The results are
\begin{align*}
A_r(\tau,\zeta) =&i\overline\gamma A_p(\tau')J_0\{2\overline\gamma\sqrt{\eta(\tau,\tau')\xi}\}H_H(\tau,\tau',\zeta,\zeta'),\tag{28a}\label{28a}\\
A_s(\tau,\zeta) =&\delta(\tau - \tau')H(\xi) - \overline\gamma A_p^*(\tau)A_p(\tau')\sqrt{\xi/\eta(\tau,\tau')}\\
\times&J_1\{2\overline\gamma\sqrt{\eta(\tau,\tau')\xi}\} H_H(\tau,\tau',\zeta,\zeta'),\tag{28b}\label{28b}
\end{align*}
\noindent where $J_n$ is a Bessel function of order $n$,
$\xi=\zeta-\zeta'$,and $H_H(\tau,\tau',\zeta,\zeta')=H(\tau -
\tau')H(\zeta-\zeta')$. In the text, these solutions are referred to
as the Green functions $G_{rs}$ and $G_{ss}$, respectively.

If an impulse is applied to the idler at the input boundary, the
transformed equations of motion and solutions are
\begin{align*}
\partial_\tau A_r(\tau,s) &= i\overline\gamma A_p(\tau)A_s(\tau,s) + \delta(\tau - \tau'),\tag{29a}\label{29a}\\
s A_s(\tau,s) &= i\overline\gamma A_p^*(\tau)A_r(\tau,s),\tag{29b}\label{29b}\\
\Rightarrow A_r(\tau,s) &= \exp[-\overline\gamma^2\eta(\tau,\tau')/s] H(\tau - \tau'),\tag{30a}\label{30a}\\
A_s(\tau,s) &= i\overline\gamma\frac{A_p^*(\tau)}{s} \exp[-\overline\gamma^2\eta(\tau,\tau')/s] H(\tau - \tau').\tag{30b}\label{30b}
\end{align*}
By using the aforementioned tables of inverse transforms, one finds that
\begin{align*}
 A_r(\tau,\zeta) =& H(\tau - \tau')\delta(\zeta-\zeta') - \overline\gamma\sqrt{\eta(\tau,\tau')/\xi}\\
\times&J_1\{2\overline\gamma\sqrt{\eta(\tau,\tau')\xi}\} H_H(\tau,\tau',\zeta,\zeta'),\tag{31a}\label{31a}\\
A_s(\tau,\zeta) = &i\overline\gamma A_p^*(\tau)J_0\{2\overline\gamma\sqrt{\eta(\tau,\tau')\xi}\} H_H(\tau,\tau',\zeta,\zeta').\tag{31b}\label{31b}
\end{align*}

These solutions are referred to as the Green functions $G_{rr}$ and
$G_{sr}$, respectively. These Green functions are stable analogs of
the Green functions for stimulated Brillouin scattering (SBS) and
stimulated Raman scattering (SRS) \cite{ray81,gia95}, and are
equivalent to the Green functions for anti-Stokes SRS \cite{ray04}.

\end{document}